\documentclass[twocolumn,showpacs,aps,prd]{revtex4-1}

\usepackage{graphicx} \graphicspath{{figures/}}
\usepackage{rotating}   
\usepackage{dcolumn} \newcolumntype{.}{D{.}{.}{-1}}               
\usepackage{atlasphysics}
\usepackage{multirow}
\usepackage{hyperref}

\newcommand{\Itemize}[1]{ \begin{itemize} #1 \end{itemize} }

\newcommand{\rec}{\ensuremath{^\mathrm{rec}}}

\newcommand{\obs}{\ensuremath{^\mathrm{obs}}}

\newcommand{\bkg}{\ensuremath{^\mathrm{bkg}}}

\newcommand{\effRA}{\ensuremath{\epsilon^{RA}}}

\newcommand{\mgg}{\ensuremath{m_\evtgamgam}}
\newcommand{\ptgg}{\ensuremath{p_{\mathrm{T},\evtgamgam}}}
\newcommand{\dphigg}{\ensuremath{\Delta\phi_\evtgamgam}}

\newcommand{\Etisol} {\ensuremath{\ET^{\mathrm{iso}}}}
\newcommand{\EtisolOne} {\ensuremath{E_\mathrm{T,1}^{\mathrm{iso}}}}
\newcommand{\EtisolTwo} {\ensuremath{E_\mathrm{T,2}^{\mathrm{iso}}}}
\newcommand{\Etpartisol} {\ensuremath{\ET^{\mathrm{iso(part)}}}}

\newcommand{\NA}   {\ensuremath{N_A}}
\newcommand{\NB}   {\ensuremath{N_B}}
\newcommand{\MA}   {\ensuremath{N_C}}
\newcommand{\MB}   {\ensuremath{N_D}}

\newcommand{\NAbkg}   {\ensuremath{\NA^{\mathrm{bkg}}}}
\newcommand{\NBbkg}   {\ensuremath{\NB^{\mathrm{bkg}}}}
\newcommand{\MAbkg}   {\ensuremath{\MA^{\mathrm{bkg}}}}
\newcommand{\MBbkg}   {\ensuremath{\MB^{\mathrm{bkg}}}}

\newcommand{\NAsig}   {\ensuremath{\NA^{\mathrm{sig}}}}

\newcommand{\NAP}   {\ensuremath{N'_A}}
\newcommand{\NBP}   {\ensuremath{N'_B}}
\newcommand{\MAP}   {\ensuremath{N'_C}}
\newcommand{\MBP}   {\ensuremath{N'_D}}

\newcommand{\NAPsig}   {\ensuremath{{\NAP}^{\mathrm{sig}}}}

\newcommand{\true}{\ensuremath{^\mathrm{true}}}

\newcommand{\loose}{\ensuremath{\mathbf{L}}}
\newcommand{\tight}{\ensuremath{\mathbf{T}}}
\newcommand{\isol}{\ensuremath{\mathbf{I}}}
\newcommand{\nontight}{\ensuremath{\mathbf{\tilde{T}}}}
\newcommand{\nonisol}{\ensuremath{\mathbf{\tilde{I}}}}

\newcommand{\evtgamgam}{\ensuremath{{\gamma\gamma}}}
\newcommand{\evtgamjet}{\ensuremath{{\gamma\mathrm{j}}}}
\newcommand{\evtjetgam}{\ensuremath{{\mathrm{j}\gamma}}}
\newcommand{\evtjetjet}{\ensuremath{{\mathrm{jj}}}}

\newcommand{\objgam}{\ensuremath{{\gamma}}}
\newcommand{\objjet}{\ensuremath{{\mathrm{j}}}}

\newcommand{\Ptag}{\ensuremath{{\hat{\mathrm{P}}}}}
\newcommand{\Ftag}{\ensuremath{{\hat{\mathrm{F}}}}}

\newcommand{\PP}{\ensuremath{{\mathrm{PP}}}}
\newcommand{\PF}{\ensuremath{{\mathrm{PF}}}}
\newcommand{\FP}{\ensuremath{{\mathrm{FP}}}}
\newcommand{\FF}{\ensuremath{{\mathrm{FF}}}}

\newcommand{\el}{\ensuremath{e}}
\newcommand{\ph}{\ensuremath{\gamma}}

\newcommand{\fph}{\ensuremath{f_{\el\rightarrow\ph}}}
\newcommand{\fel}{\ensuremath{f_{\ph\rightarrow\el}}}

\newcommand{\de}{\mbox{d}}

\begin{document}

%
%
\begin{minipage}[t]{\textwidth}
  \begin{flushright}
    CERN-PH-EP-2011-088\\
    Submitted to Physical Review D
    \end{flushright}
\end{minipage}
\vspace{0.5cm}

\title{\boldmath Measurement of the isolated diphoton cross-section \\ 
  in $pp$ collisions at $\sqrt{s} = 7 \TeV$ with the ATLAS detector}

\author{G. Aad \textit{et al.}}\thanks{Full author list given at the end of the article.}
\collaboration{The ATLAS Collaboration}

\date{September 26, 2011}

\begin{abstract}
  The ATLAS experiment has measured the production cross-section of
  events with two isolated photons in the final state, in
  proton-proton collisions at $\sqrt{s} = 7~\tev$.  The full data set
  acquired in 2010 is used, corresponding to an integrated luminosity of
  $37~\ipb$. The background, consisting of
  hadronic jets and isolated electrons, is estimated with fully
  data-driven techniques and subtracted. The differential
  cross-sections, as functions of the di-photon mass (\mgg), total transverse
  momentum (\ptgg) and azimuthal separation (\dphigg), are presented and
  compared to the predictions of next-to-leading-order QCD.
\end{abstract}


\maketitle


\newpage

\section{Introduction}
\label{sec:Introduction}

The production of di-photon final states in proton-proton collisions
may occur through quark-antiquark $t$-channel annihilation,
$q\bar{q}\rightarrow\gamma\gamma$, or via gluon-gluon interactions,
$gg\rightarrow\gamma\gamma$, mediated by a quark box diagram.  Despite
the higher order of the latter, the two contributions are comparable,
due to the large gluon flux at the LHC.  Photon-parton production with
photon radiation also contributes in processes such as
$q\bar{q},gg\rightarrow g\gamma\gamma$ and $qg\rightarrow
q\gamma\gamma$. During the parton fragmentation process, more photons
may also be produced. In this analysis, all such photons are
considered as signal if they are isolated from other activity in the
event. Photons produced after the hadronization by neutral hadron
decays, or coming from radiative decays of other particles, are
considered as part of the background.

The measurement of the di-photon production cross-section at the LHC
is of great interest as a probe of QCD, especially in some particular
kinematic regions.  For instance, the distribution of the azimuthal
separation, \dphigg, is sensitive to the fragmentation model,
especially when both photons originate from fragmentation.  On the
other hand, for balanced back-to-back di-photons ($\dphigg\simeq\pi$
and small total transverse momentum, \ptgg) the production is
sensitive to soft gluon emission, which is not accurately described by
fixed-order perturbation theory.

Di-photon production is also an irreducible background for some new
physics processes, such as the Higgs decay into photon pairs
\cite{ProcLHCWS1990}: in this case, the spectrum of the invariant
mass, \mgg, of the pair is analysed, searching for a resonance.
Moreover, di-photon production is a characteristic signature of some
exotic models beyond the Standard Model. For instance, Universal
Extra-Dimensions (UED) predict non-resonant di-photon production
associated with significant missing transverse energy
\cite{UED,PhysRevLett.106.121803}. Other extra-dimension models, such
as Randall-Sundrum \cite{PhysRevLett.83.3370}, predict the production
of gravitons, which would decay into photon pairs with a narrow width.

Recent cross-section measurements of di-photon production at hadron
colliders have been performed by the D\O~\cite{D0_diphoton} and
CDF~\cite{CDF_diphoton_2011} collaborations, at the Tevatron
proton-antiproton collider with a centre-of-mass energy
$\sqrt{s}=1.96$~TeV.

In this document, di-photon production is studied in proton-proton
collisions at the LHC, with a centre-of-mass energy
$\sqrt{s}=7$~TeV. After a short description of the ATLAS detector
(Section~\ref{sec:Detector}), the analysed collision data and the
event selection are detailed in Section~\ref{sec:Selection}, while the
supporting simulation samples are listed in
Section~\ref{sec:MCSamples}.  The isolation properties of the signal
and of the hadronic background are studied in
Section~\ref{sec:isolation}.  The evaluation of the di-photon signal
yield is obtained by subtracting the backgrounds from hadronic jets
and from isolated electrons, estimated with data-driven methods as
explained in Section~\ref{sec:Backgrounds}.
Section~\ref{sec:efficiency} describes how the event selection
efficiency is evaluated and how the final yield is obtained.  Finally,
in Section~\ref{sec:cross-section}, the differential cross-section of
di-photon production is presented as a function of \mgg, \ptgg\ and
\dphigg.

\section{The ATLAS detector}
\label{sec:Detector}

The ATLAS detector~\cite{ATLAS_detector} is a multipurpose particle
physics apparatus with a forward-backward symmetric cylindrical
geometry and near $4\pi$ coverage in solid angle.  ATLAS uses a
right-handed coordinate system with its origin at the nominal
interaction point (IP) in the centre of the detector and the $z$-axis
along the beam pipe. The $x$-axis points from the IP to the centre of
the LHC ring, and the $y$ axis points upward. Cylindrical coordinates
$(r,\phi)$ are used in the transverse plane, $\phi$ being the
azimuthal angle around the beam pipe. The pseudorapidity is defined in
terms of the polar angle $\theta$ as $\eta=-\ln\tan(\theta/2)$.  The
transverse momentum is defined as $\pt=p\sin\theta=p/\cosh\eta$, and a
similar definition holds for the transverse energy \et.

The inner tracking detector (ID) covers the pseudorapidity range
$|\eta|<2.5$, and consists of a silicon pixel detector, a silicon
microstrip detector, and a transition radiation tracker in the range
$|\eta|<2.0$.  The ID is surrounded by a superconducting solenoid
providing a 2~T magnetic field.  The inner detector allows an accurate
reconstruction of tracks from the primary proton-proton collision
region, and also identifies tracks from secondary vertices, permitting
the efficient reconstruction of photon conversions in the inner
detector up to a radius of $\approx$~80~cm.

The electromagnetic calorimeter (ECAL) is a lead-liquid argon (LAr)
sampling calorimeter with an accordion geometry.  It is divided into a
barrel section, covering the pseudorapidity region $|\eta|< 1.475$,
and two end-cap sections, covering the pseudorapidity regions
$1.375<|\eta|<3.2$.  It consists of three longitudinal layers. The
first one, in the ranges $|\eta|<1.4$ and $1.5<|\eta|<2.4$, is
segmented into high granularity ``strips'' in the $\eta$ direction,
sufficient to provide an event-by-event discrimination between single
photon showers and two overlapping showers coming from a $\pi^0$
decay.  The second layer of the electromagnetic calorimeter, which
collects most of the energy deposited in the calorimeter by the photon
shower, has a thickness of about 17 radiation lengths and a
granularity of $0.025\times0.025$ in $\eta\times\phi$ (corresponding
to one cell).  A third layer is used to correct leakage beyond the
ECAL for high energy showers. In front of the accordion calorimeter a
thin presampler layer, covering the pseudorapidity interval
$|\eta|<1.8$, is used to correct for energy loss before the
calorimeter.

The hadronic calorimeter (HCAL), surrounding the ECAL, consists of an
iron-scintillator tile calorimeter in the range $|\eta|<1.7$, and two
copper-LAr calorimeters spanning $1.5<|\eta|<3.2$. The acceptance is
extended by two tungsten-LAr forward calorimeters up to $|\eta|<4.9$.
The muon spectrometer, located beyond the calorimeters, consists of
three large air-core superconducting toroid systems, precision
tracking chambers providing accurate muon tracking over $|\eta|<2.7$,
and fast detectors for triggering over $|\eta|<2.4$.

A three-level trigger system is used to select events containing two
photon candidates.  The first level trigger (level-1) is hardware
based: using a coarser cell granularity ($0.1\times 0.1$ in
$\eta\times\phi$), it searches for electromagnetic deposits with a
transverse energy above a programmable threshold.  The second and
third level triggers (collectively referred to as the ``high-level''
trigger) are implemented in software and exploit the full granularity
and energy calibration of the calorimeter.

\section{Collision data and selections}
\label{sec:Selection}

The analysed data set consists of proton-proton collision data at
$\sqrt{s}=7$~TeV collected in 2010, corresponding to an integrated
luminosity of $37.2\pm1.3~\ipb$~\cite{ATLAS_LUMI_all}.  The events are
considered only when the beam condition is stable and the trigger
system, the tracking devices and the calorimeters are operational.

\subsection{Photon reconstruction}
\label{sec:PhotonReconstruction}

A photon is defined starting from a cluster in the ECAL.  
If there are no tracks pointing to
the cluster, the object is classified as an {\em unconverted photon}.
In case of {\em converted photons}, one or two tracks may be
associated to the cluster, therefore creating an ambiguity in the
classification with respect to electrons. This is addressed as
described in Ref~\cite{ATL-PHYS-PUB-2011-007}.

A fiducial acceptance is required in pseudorapidity,
$|\eta^\gamma|<2.37$, with the exclusion of the barrel/endcap
transition $1.37<|\eta^\gamma|<1.52$.  This corresponds to the regions
where the ECAL ``strips'' granularity is more effective for photon identification and jet
background rejection~\cite{ATL-PHYS-PUB-2011-007}.  Moreover, photons
reconstructed near to regions affected by read-out or high-voltage
failures are not considered.

In the considered acceptance range, the uncertainty on the photon
energy scale is estimated to be $\sim\pm1\%$.  The energy resolution is
parametrized as
$\sigma_E/E \simeq a/\sqrt{E~{\rm [GeV]}} \oplus c$, where the
sampling term $a$ varies between 10\% and 20\% depending on
$\eta^\gamma$, and the constant term $c$ is estimated to be 1.1\% in
the barrel and 1.8\% in the endcap. 
Such a performance has been measured in $Z\rightarrow e^+e^-$ events observed in proton-proton collision data in 2010.

\subsection{Photon selection}
\label{sec:PhotonSelection}

The photon sample suffers from a major background due to hadronic jets, which generally produce calorimetric deposits broader and
less isolated than electromagnetic showers, with sizable energy leaking to the HCAL. 
Most of the background is reduced by applying requirements (referred to as the \textsc{loose} selection, \loose) 
on the energy fraction measured in the HCAL, and on the shower width measured by the 
second layer of the ECAL. The remaining background is mostly due to photon pairs from neutral hadron decays
(mainly $\pi^0$) with a small opening angle, and reconstructed as single photons.
This background is further reduced by applying a more stringent selection on the shower width in the second ECAL layer, together with 
additional requirements on the shower shape measured by the first ECAL layer: a narrow shower width and the absence of a second significant maximum in the energy 
deposited in contiguous strips. The combination of all these requirements is referred to as the \textsc{tight} selection (\tight). 
Since converted photons tend to have broader shower shapes than unconverted ones, the cuts of the \textsc{tight} selection are tuned differently for the two
photon categories.
More details on these selection criteria are given in Ref~\cite{ATLAS-INCLPHOTON}.

To reduce the jet background further, an isolation requirement is applied: the isolation transverse energy \Etisol{}, measured by the
calorimeters in a cone of angular radius $R=\sqrt{(\eta-\eta^\gamma)^2+(\phi-\phi^\gamma)^2}<0.4$, is required to satisfy $\Etisol<3$~GeV (isolated photon, \isol). 
The calculation of \Etisol{}
is performed summing over ECAL and HCAL cells
surrounding the photon candidate, after removing a central core that contains most of the photon energy. An out-of-core energy correction~\cite{ATLAS-INCLPHOTON} 
is applied, to make
\Etisol\ essentially independent of $\et^\gamma$, and an ambient energy correction, based on the measurement of soft 
jets~\cite{Cacciari:UE,Cacciari:area} 
is applied on an event-by-event basis, to
remove the contribution from the underlying event and from additional proton-proton interactions (``in-time pile-up'').

\subsection{Event selection}
\label{sec:EventSelection}

The di-photon candidate events are selected according to the following steps:
\Itemize{
\item The events are selected by a di-photon trigger, in which both photon candidates must satisfy the trigger 
selection and have a transverse energy
\mbox{$\et^\gamma>15$~GeV}. To select genuine collisions, at least one primary vertex with three or more tracks must be reconstructed. 
\item The event must contain at least
two photon candidates, with $\et^\gamma>16$~GeV, in the acceptance defined in Section~\ref{sec:PhotonReconstruction}, and passing the \textsc{loose} selection. 
If more than two such photons exist, the two with highest $\et^\gamma$ are chosen. 
\item To avoid a too large overlap between the two isolation cones, an angular
separation $\Delta R_\evtgamgam=\sqrt{(\eta_1^\gamma-\eta_2^\gamma)^2+(\phi_1^\gamma-\phi_2^\gamma)^2}>0.4$ is required. 
\item Both photons must satisfy the \textsc{tight} selection (\tight\tight{} sample).
\item Both photons must satisfy the isolation requirement $\Etisol<3$~GeV (\tight\isol\tight\isol{} sample).
}
In the analysed data set, there are 63673 events where both photons satisfy the \textsc{loose} selection and the $\Delta R_\evtgamgam$ separation requirement.
Among these, 5365 events belong to the \tight\tight{} sample, and 2022 to the \tight\isol\tight\isol{} sample.

\section{Simulated events}
\label{sec:MCSamples}

The characteristics of the signal and background events are
investigated with Monte~Carlo samples, generated using
\textsc{Pythia}~6.4.21~\cite{pythia}.  The simulated samples are
generated with pile-up conditions similar to those under which most of
the data were taken.  Particle interactions with the detector
materials are modelled with \textsc{Geant4}~\cite{geant} and the
detector response is simulated.  The events are reconstructed with the
same algorithms used for collision data. More details on the event
generation and simulation infrastructure are provided in
Ref~\cite{ATLAS_simulation}.

The di-photon signal is generated with \textsc{Pythia}, where photons
from both hard scattering and quark bremsstrahlung are modelled.  To
study systematic effects due to the generator model, an alternative
di-photon sample has been produced with \textsc{Sherpa}~\cite{sherpa}.

The background processes are generated with the main
physical processes that produce (at least) two sizable calorimetric
deposits: these include di-jet and photon-jet final states, but minor
contributions, e.g. from $W,Z$ bosons, 
are also present. 
Such a Monte~Carlo sample, referred to as ``di-jet-like'', provides a
realistic mixture of the main final states expected to contribute to
the selected data sample.
Moreover, dedicated samples of $W\rightarrow e\nu$ and $Z\rightarrow
e^+e^-$ simulated events are used for the electron/photon comparison
in isolation and background studies.

\section{Properties of the isolation transverse energy}
\label{sec:isolation}

The isolation transverse energy, \Etisol, is a powerful discriminating
variable to estimate the jet background contamination in the sample of photon
candidates.  The advantage of using this quantity is that its
distribution can be extracted directly from the observed collision
data, both for the signal and the background, without relying
on simulations.

Section~\ref{sec:isolation-from-photons} describes a method to extract
the distribution of \Etisol\ for background and signal, from observed
photon candidates.  An independent method to extract the signal
\Etisol\ distribution, based on observed electrons, is described in
Section~\ref{sec:isolation-from-electrons}.  Finally, the correlation
between isolation energies in events with two photon candidates is
discussed in Section~\ref{sec:isolation-two-candidates}.

\subsection{Background and signal isolation from photon candidates}
\label{sec:isolation-from-photons}

\begin{figure}
  \centering
  \includegraphics[width=\columnwidth]{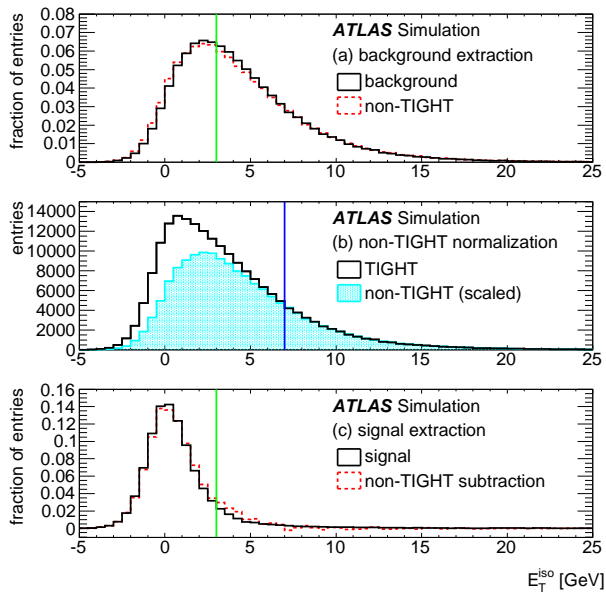}
  \caption{
    Extraction of the isolation energy (\Etisol) distributions, for signal and background. 
    The plots are made with a ``di-jet-like'' Monte~Carlo sample: 
    the ``signal'' and ``background'' classifications are based on the Monte~Carlo information. 
    (a) Normalized \Etisol{} distribution for the background and for the non-\textsc{tight} sample. 
    (b) \Etisol{} distribution, for the \textsc{tight} and the non-\textsc{tight} samples: 
    the latter is scaled as explained in the text. 
    (c) Normalized \Etisol{} distribution for the signal and for the \textsc{tight} sample, after subtracting the 
    scaled non-\textsc{tight} sample. 
    In (a,c) the vertical line shows the isolation cut $\Etisol<3$~GeV.
  }
  \label{fig:4x4-dd-eff-fake}
\end{figure}

For the background study, a control sample is defined by reconstructed
photons that fail the \textsc{tight} selection but pass a looser one,
where some cuts are released on the shower shapes measured by the ECAL
``strips''.  Such photons are referred to as non-\textsc{tight}. A
study carried out on the ``di-jet-like'' Monte~Carlo sample shows that
the \Etisol\ distribution in the non-\textsc{tight} sample reproduces
that of the background, as shown in
Figure~\ref{fig:4x4-dd-eff-fake}(a).

The \textsc{tight} photon sample contains a mixture of signal and
background. However, a comparison between the shapes of the \Etisol{}
distributions from \textsc{tight} and non-\textsc{tight} samples
(Figure~\ref{fig:4x4-dd-eff-fake}(b)) shows that for $\Etisol>7$~GeV
there is essentially no signal in the \textsc{tight} sample.
Therefore, the background contamination in the \textsc{tight} sample
can be subtracted by using the non-\textsc{tight} sample, normalized
such that the integrals of the two distributions are equal for
$\Etisol>7$~GeV. The \Etisol{} distribution of the signal alone is
thus extracted.  Figure~\ref{fig:4x4-dd-eff-fake}(c) shows the result,
for photons in the ``di-jet-like'' Monte~Carlo sample.

In collision data, events with two photon candidates are used to build
the \textsc{tight} and non-\textsc{tight} samples, for the leading and
subleading candidate separately. The points in
Figure~\ref{fig:signalIsolPlot} display the distribution of \Etisol\
for the leading and sub-leading photons.  In each of the two
distributions, one bin has higher content, reflecting opposing
fluctuations in the subtracted input distributions in those bins. The
effect on the di-photon cross-section measurement is negligible.

The main source of systematic error comes from the definition of the
non-\textsc{tight} control sample.  There are three sets of strips
cuts that could be released: the first set concerns the shower width
in the core, the second tests for the presence of two maxima in the
cluster, and the third is a cut on the full shower width in the
strips. The choice adopted is to release only the first two sets of
cuts, as the best compromise between maximizing the statistics in the
control sample, while keeping the background \Etisol{} distribution
fairly unbiased.  To test the effect of this choice, the sets of
released cuts have been changed, either by releasing only the cuts on
the shower core width in the strips, or by releasing all the strips
cuts.  A minor effect is also due to the choice of the region
$\Etisol>7$~GeV, to normalize the non-\textsc{tight} control sample:
the cut has therefore been moved to 6 and 8~GeV.

More studies with the ``di-jet-like'' Monte~Carlo sample have been
performed, to test the robustness of the \Etisol\ extraction against
model-dependent effects such as: ({\em i}) signal leakage into the
non-\textsc{tight} sample; ({\em ii}) correlations between \Etisol{}
and strips cuts; ({\em iii}) different signal composition, i.e.
fraction of photons produced by the hard scattering or by the
fragmentation process; ({\em iv}) different background composition,
i.e. fraction of photon pairs from $\pi^0$ decays. In all cases, the
overall systematic error, computed as described above, covers the
differences between the true and data-driven results as evaluated from
these Monte~Carlo tests.

\subsection{Signal isolation from electron extrapolation}
\label{sec:isolation-from-electrons}

\begin{figure}
  \centering
  \includegraphics[width=\columnwidth]{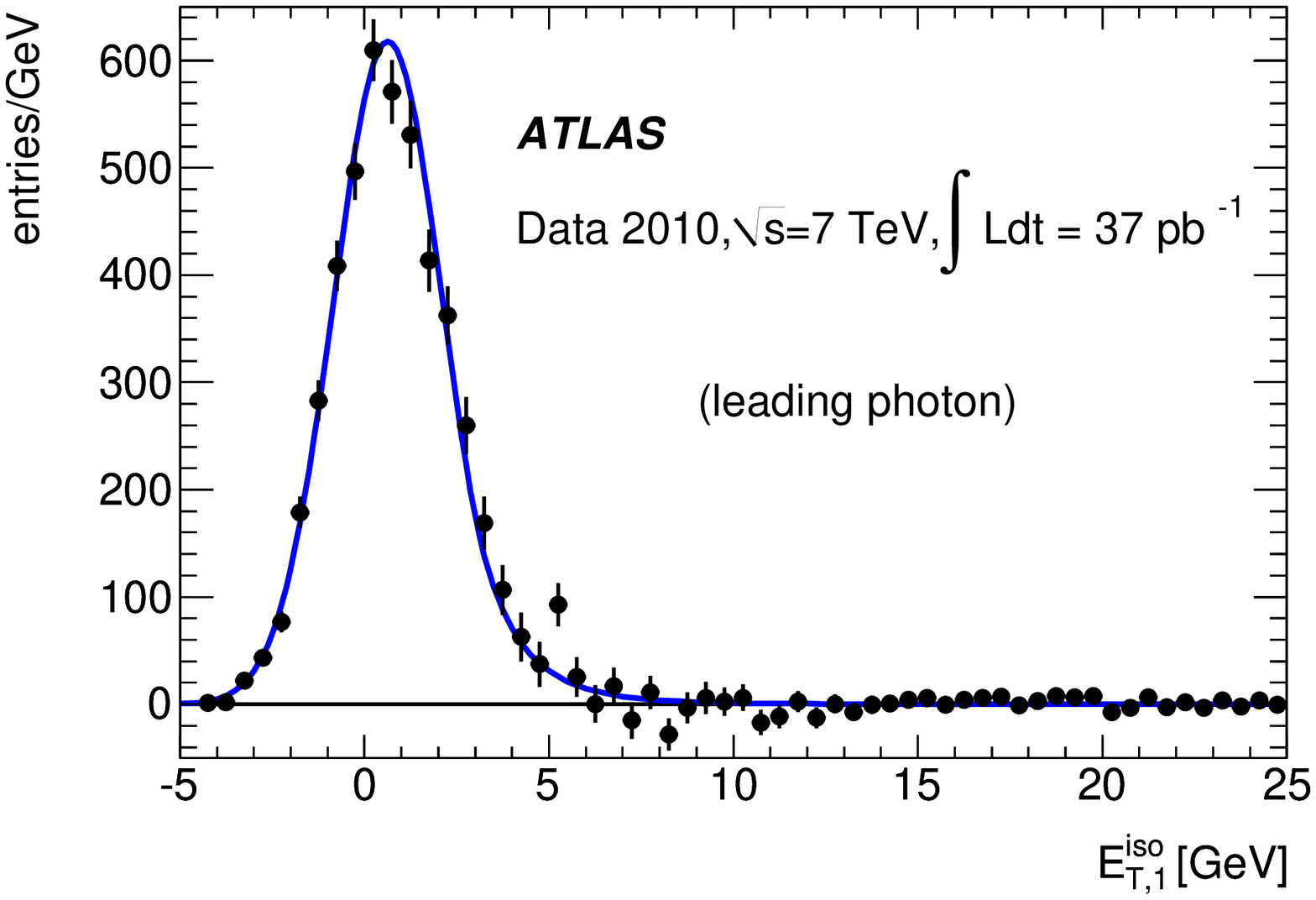}
  \includegraphics[width=\columnwidth]{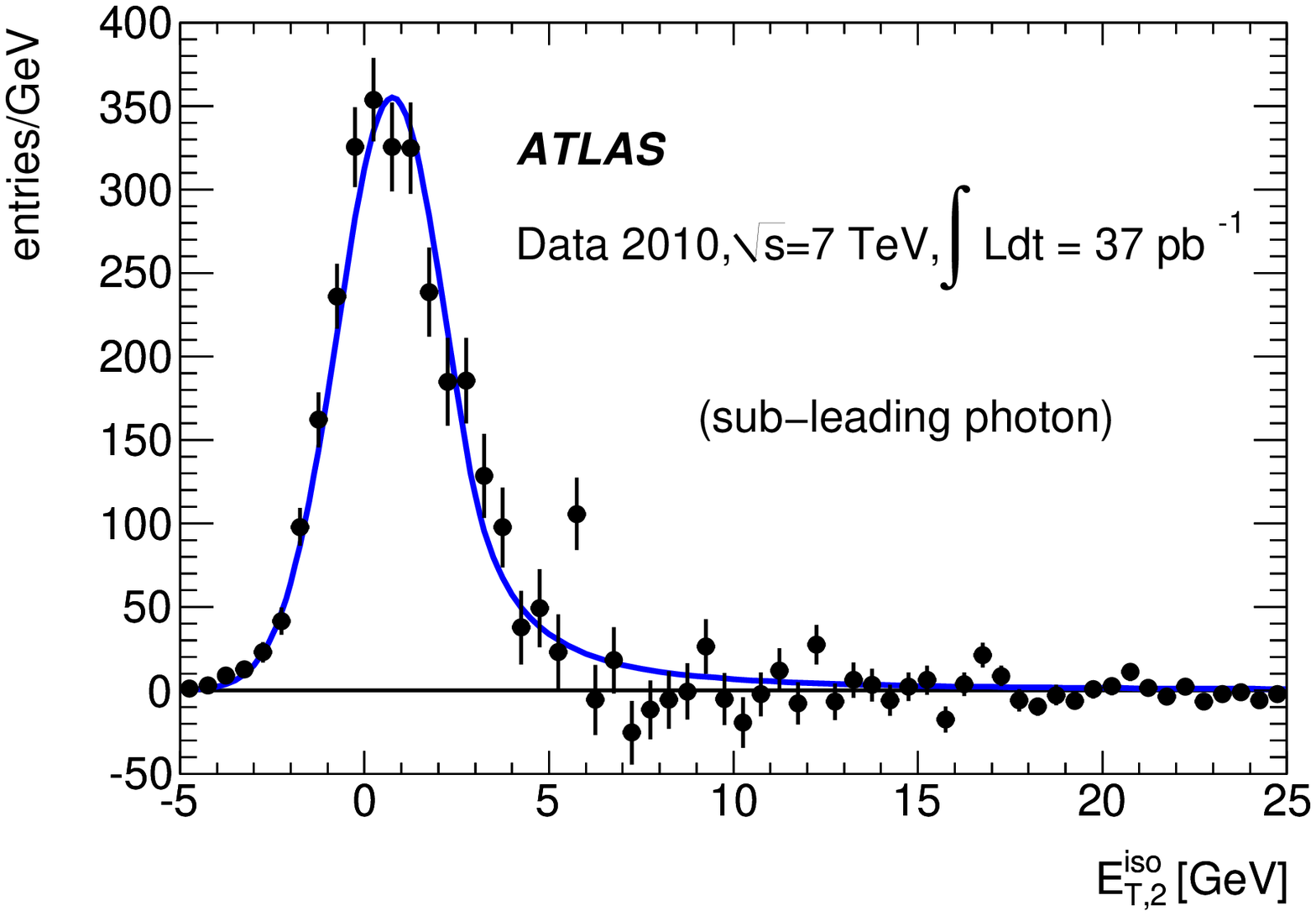}
  \caption{Data-driven signal isolation distributions for the leading (top) and sub-leading (bottom)
    photons obtained using the photon candidates (solid circles) or extrapolated from electrons (continuous lines).
  }
  \label{fig:signalIsolPlot}
\end{figure}

An independent method of extracting the \Etisol{} distribution for the
signal photons is provided by the ``electron extrapolation''. In
contrast to photons, it is easy to select a pure electron sample from
data, from $W^\pm\rightarrow e^\pm\nu$ and $Z\rightarrow e^+e^-$
events~\cite{ATLAS_WZcrosssection_2010}.  The main differences between
the electron and photon \Etisol{} distributions are: ({\em i}) the
electron \Etisol{} in the bulk of the distribution is slightly larger,
because of bremsstrahlung in the material upstream of the calorimeter;
({\em ii}) the photon \Etisol{} distribution exhibits a larger tail
because of the contribution of the photons from fragmentation,
especially for the sub-leading photon.  Such differences are
quantified with $W^\pm\rightarrow e^\pm\nu$, $Z\rightarrow e^+e^-$ and
\evtgamgam{} Monte~Carlo samples by fitting the \Etisol{}
distributions with Crystal~Ball functions~\cite{CrystalBall} and
comparing the parameters. Then, the electron/photon differences are
propagated to the selected electrons from collision data.  The result
is shown by the continuous lines in Figure~\ref{fig:signalIsolPlot},
agreeing well with the \Etisol{} distributions obtained from the
non-\textsc{tight} sample subtraction (circles).

\subsection{Signal and background isolation in events with two photon candidates}
\label{sec:isolation-two-candidates}

In events with two photon candidates, possible correlations between
the two isolation energies have been investigated by studying the
signal and background \Etisol{} distributions of a candidate
(``probe'') under different isolation conditions of the other
candidate (``tag'').  The signal \Etisol{} shows negligible dependence
on the tag conditions. In contrast, the background \Etisol{}
exhibits a clear positive correlation with the isolation transverse
energy of the tag: if the tag passes (or fails) the isolation
requirement, the probe background candidate is more (or less)
isolated. This effect is visible especially in di-jet final states,
which can be directly studied in collision data by requiring both
photon candidates to be non-\textsc{tight}, and is taken into account in the jet background
estimation (Section~\ref{sec:Jet}).

This correlation is also visible in the ``di-jet-like'' Monte~Carlo sample.

\section{Background subtraction and signal yield determination}
\label{sec:Backgrounds}

The main background to selected photon candidates consists of hadronic
jets. This is reduced by the photon \textsc{tight} selection described
in Section~\ref{sec:PhotonSelection}. However a significant component
is still present and must be subtracted. The techniques to achieve
this are described in Section~\ref{sec:Jet}.

Another sizable background component comes from isolated electrons,
mainly originating from $W$ and $Z$ decays, which look similar to
photons from the calorimetric point of view. The subtraction of such a
contamination is addressed in Section~\ref{sec:ElectronBkd}.

The background due to cosmic rays and to beam-gas collisions has been
studied on dedicated data sets, selected by special triggers. Its
impact is found to be negligible.

\subsection{Jet background}
\label{sec:Jet}

\begin{figure}
  \centering
  \includegraphics[width=.9\columnwidth]{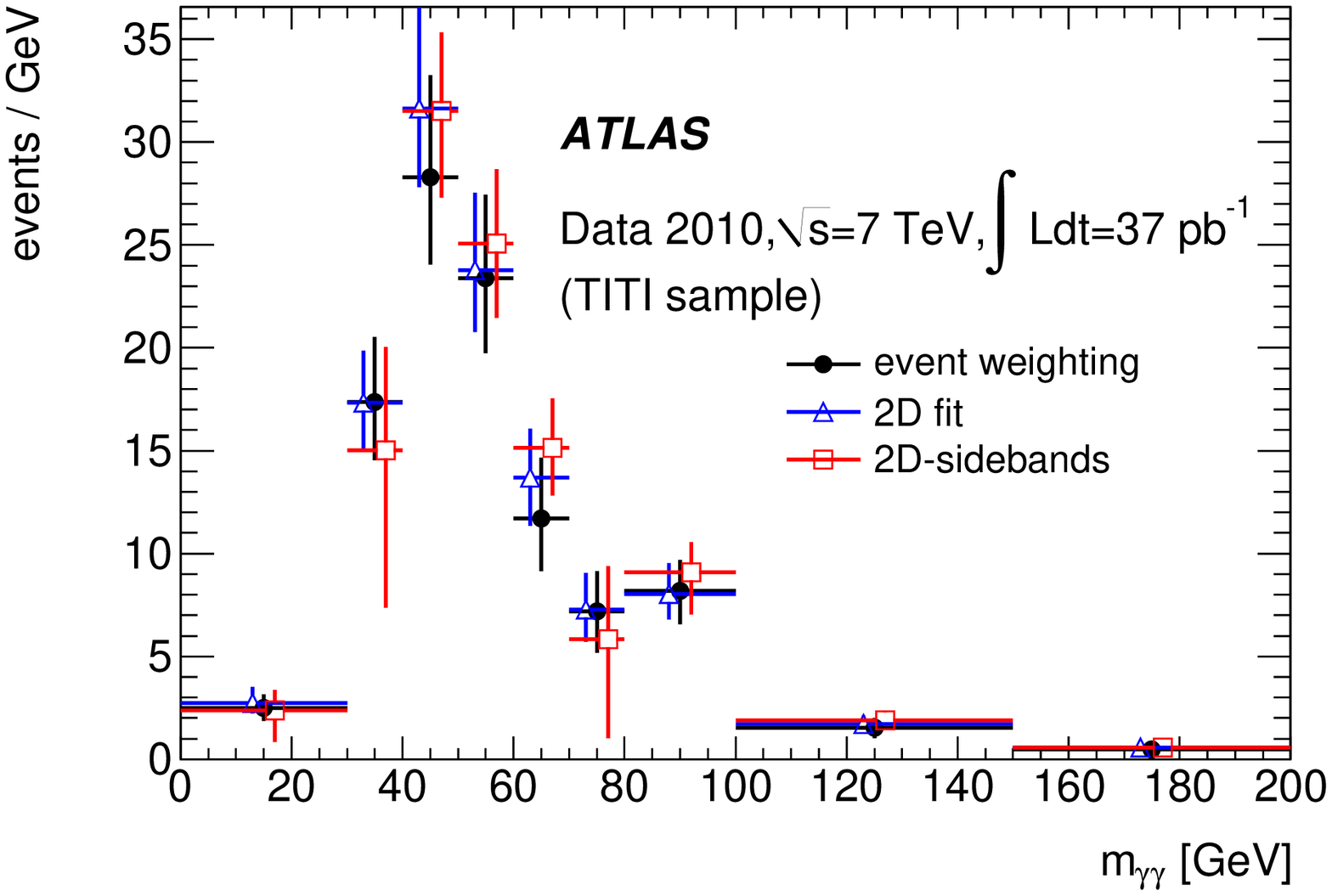}
  \includegraphics[width=.9\columnwidth]{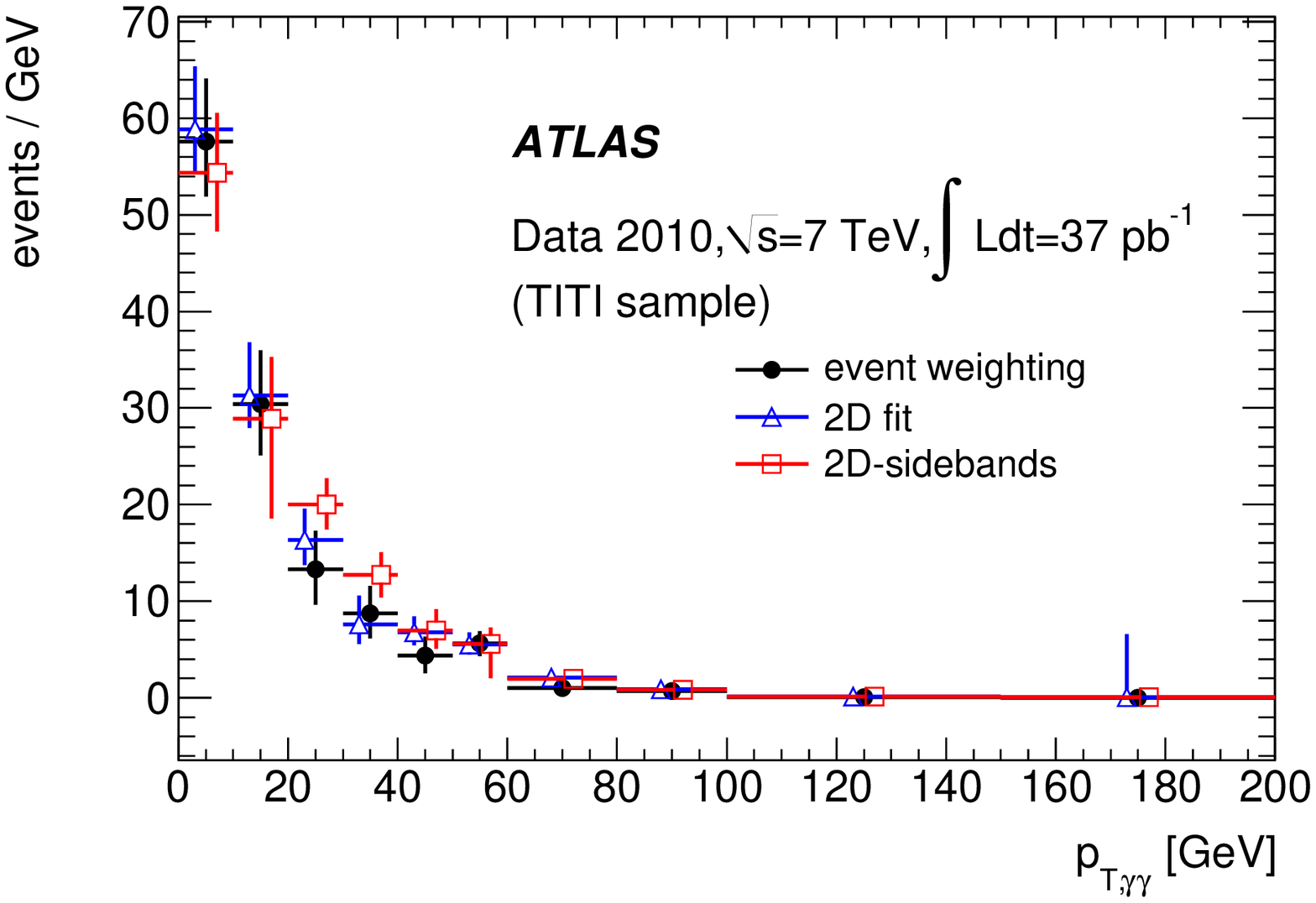}
  \includegraphics[width=.9\columnwidth]{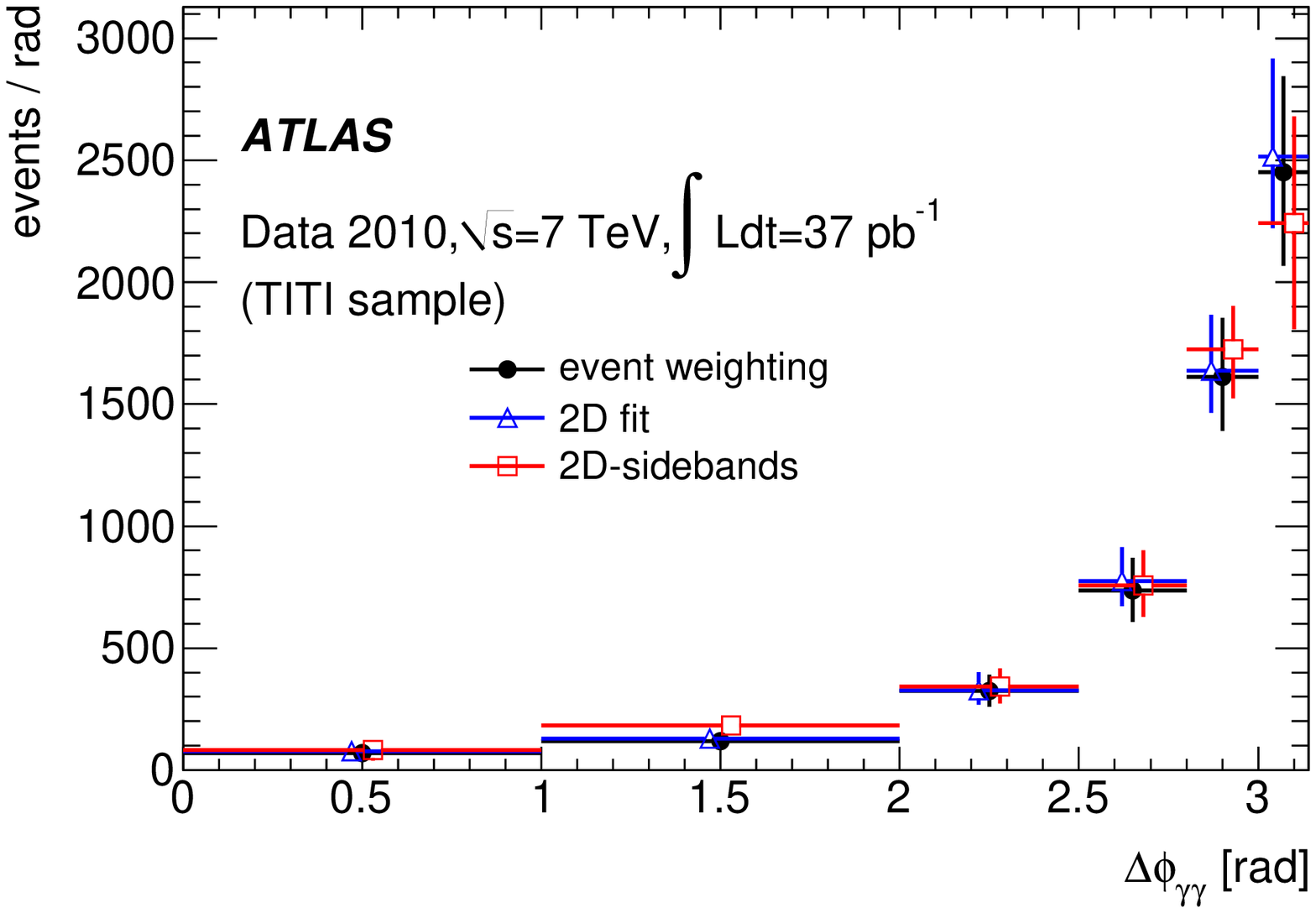}
  \caption{ Differential \evtgamgam\ yields in the
    \tight\isol\tight\isol\ sample
    ($N^{\tight\isol\tight\isol}_\evtgamgam$), as a function of the
    three observables \mgg, \ptgg, \dphigg, obtained with the three
    methods. In each bin, the yield is divided by the bin width.  The
    vertical error bars display the total errors, accounting for both
    the statistical uncertainties and the systematic effects.  The
    points are artificially shifted horizontally, to better display
    the three results.  }
  \label{fig:comparison}
\end{figure}

The jet background is due to photon-jet and di-jet final states. This
section describes three methods, all based on the isolation transverse
energy, \Etisol, aiming to separate the \tight\isol\tight\isol{}
sample into four categories:
$$ N^{\tight\isol\tight\isol}_\evtgamgam ~,~ N^{\tight\isol\tight\isol}_\evtgamjet ~,~ N^{\tight\isol\tight\isol}_\evtjetgam ~,~ N^{\tight\isol\tight\isol}_\evtjetjet ~,~ $$
according to their physical final states --- \evtgamjet{} and
\evtjetgam{} differ by the jet faking respectively the sub-leading or
the leading photon candidate.  The signal yield
$N^{\tight\isol\tight\isol}_\evtgamgam$ is evaluated in bins of the
three observables \mgg, \ptgg, \dphigg, as in
Figure~\ref{fig:comparison}.  Due to the dominant back-to-back
topology of di-photon events, the kinematic selection produces a
turn-on in the distribution of the di-photon invariant mass, at
$\mgg\gtrsim2\et^\mathrm{cut}$ ($\et^\mathrm{cut}=16$~GeV being the
applied cut on the photon transverse energy), followed by the usual
decrease typical of the continuum processes. The region at lower \mgg\
is populated by di-photon events with low $\Delta\phi$.

The excess in the mass bin $80<\mgg<100$~GeV, due to a contamination
of electrons from $Z$-decays, is addressed in
Section~\ref{sec:ElectronBkd}.

From the evaluation of the background yields
($N^{\tight\isol\tight\isol}_\evtgamjet+N^{\tight\isol\tight\isol}_\evtjetgam$
and $N^{\tight\isol\tight\isol}_\evtjetjet$), the average fractions of
photon-jet and di-jet events in the \tight\isol\tight\isol{} sample
are $\sim26\%$ and $\sim9\%$ respectively.

The three results shown in Figure~\ref{fig:comparison} are compatible.
This suggests that there are no hidden biases induced by the analyses.
However, the three measures cannot be combined, as all make use of the
same quantities --- \Etisol{} and shower shapes --- and use the
non-\textsc{tight} background control region, so they may have
correlations. None of the methods has striking advantages with respect
to the others, and the systematic uncertainties are comparable.  The
``event weighting'' method (\ref{sec:4x4Matrix}) is used for the
cross-section evaluation, since it provides event weights that are
also useful in the event efficiency evaluation, and its sources of
systematic uncertainties are independent of those related to the
signal modelling and reconstruction.

\subsubsection{Event weighting}
\label{sec:4x4Matrix}

Each event satisfying the \textsc{tight} selection on both photons
(sample \tight\tight) is classified according to whether the photons
pass or fail the isolation requirement, resulting in a \PP, \PF, \FP,
or \FF\ classification.  These are translated into four event weights
$W_\evtgamgam,W_\evtgamjet,W_\evtjetgam,W_\evtjetjet$, which describe
how likely the event is to belong to each of the four final states.  A
similar approach has already been used by the D\O~\cite{D0_diphoton}
and CDF~\cite{CDF_diphoton_2011} collaborations.

The connection between the pass/fail outcome and the weights, for the $k$-th event, is:
\begin{equation}
  \left(\begin{array}{c} S^{(k)}_\PP \\ S^{(k)}_\PF \\ S^{(k)}_\FP \\ S^{(k)}_\FF \end{array}\right) = \mathbf{E}^{(k)}
  \left(\begin{array}{c} W^{(k)}_\evtgamgam \\ W^{(k)}_\evtgamjet \\ W^{(k)}_\evtjetgam \\ W^{(k)}_\evtjetjet\end{array}\right)
  ~.
  \label{eq:4x4equation}
\end{equation}
If applied to a large number of events, the quantities $S_{XY}$ would
be the fractions of events satisfying each pass/fail classification,
and the weights would be the fractions of events belonging to the four
different final states.  On an event-by-event approach, $S^{(k)}_{XY}$
are boolean status variables (e.g. for an event where both candidates
are isolated, $S^{(k)}_\PP=1$ and
$S^{(k)}_\PF=S^{(k)}_\FP=S^{(k)}_\FF=0$).  The quantity
$\mathbf{E}^{(k)}$ is a 4$\times$4 matrix, whose coefficients give the
probability that a given final state produces a certain pass/fail
status.  If there were no correlation between the isolation transverse
energies of the two candidates, it would have the form:
\begin{equation}{\scriptsize\hspace{-.02\columnwidth}
    \left(\hspace{-.02\columnwidth}\begin{array}{cccc}
        \\
        \epsilon_1\epsilon_2         & \epsilon_1f_2         & f_1\epsilon_2         & f_1 f_2 \\
        \epsilon_1(1-\epsilon_2)     & \epsilon_1(1-f_2)     & f_1(1-\epsilon_2)     & f_1(1-f_2) \\
        (1-\epsilon_1)\epsilon_2     & (1-\epsilon_1)f_2     & (1-f_1)\epsilon_2     & (1-f_1)f_2 \\
        (1-\epsilon_1)(1-\epsilon_2) & (1-\epsilon_1)(1-f_2) & (1-f_1)(1-\epsilon_2) & (1-f_1)(1-f_2) \\
        \\
      \end{array}\hspace{-.02\columnwidth}\right) 
  }
  \label{eq:4x4equation-nocorr}
\end{equation}
where $\epsilon_i$ and $f_i$ ($i=1,2$ for the leading/sub-leading
candidate) are the probabilities that a signal or a fake photon
respectively pass the isolation cut. These are obtained from the
\Etisol{} distributions extracted from collision data, as described in
Section~\ref{sec:isolation-from-photons}.
The value of $\epsilon$ is essentially independent of $\et^\gamma$ and
changes with $\eta^\gamma$, ranging between 80\% and 95\%.  The value
of $f$ depends on both $\et$ and $\eta$ and takes values between 20\%
and 40\%. Given such dependence on the kinematics, the matrix
$\mathbf{E}^{(k)}$ is also evaluated for each event.

Due to the presence of correlation, the matrix coefficients in
equation~(\ref{eq:4x4equation-nocorr}) actually involve conditional
probabilities, depending on the pass/fail status of the other
candidate (tag) of the pair. For instance, the first two coefficients
in the last column become:
\begin{eqnarray*}
  f_1 f_2 &\rightarrow& \frac{1}{2}\left[ f_1^\Ptag f_2 + f_1 f_2^\Ptag \right] ~,\\
  f_1(1-f_2) &\rightarrow& \frac{1}{2}\left[ f_1^\Ftag(1-f_2) + f_1(1-f_2^\Ptag) \right] ~,
\end{eqnarray*}
where the superscripts $\Ptag$ and $\Ftag$ denote the pass/fail status
of the tag.  The ambiguity in the choice of the tag is solved by
taking both choices and averaging them. All the conditional
($\epsilon_i^{\Ptag,\Ftag},f_i^{\Ptag,\Ftag}$) probabilities are
derived from collision data, as discussed in
Section~\ref{sec:isolation-two-candidates}.

The signal
yield 
in the \tight\isol\tight\isol{} sample can be computed as a sum of
weights running over all events in the \tight\tight{} sample:
\begin{equation}
  N_{\evtgamgam}^{\tight\isol\tight\isol} = \sum_{k\in\tight\tight}w^{(k)} \pm \sqrt{\sum_{k\in\tight\tight}\left[w^{(k)}\right]^2} 
  ~,
  \label{eq:4x4-yield-TITI}
\end{equation}
where the weight $w^{(k)}$ for the $k$-th event is:
\begin{equation}
  w^{(k)} =  W_\evtgamgam^{(k)} ~\epsilon_1^{(k)}\epsilon_2^{(k)} 
  \label{eq:4x4-weight-TITI}
\end{equation}
and the sum over $k$ is carried out on the events in a given bin of
the variable of interest (\mgg, \ptgg, \dphigg).  The result is shown
in Figure~\ref{fig:comparison}, by the solid circles.

The main sources of systematic errors are: ({\em i}) the definition of
the non-\textsc{tight} control sample: ${}^{+12\%}_{-9\%}$; ({\em ii})
the normalization of the non-\textsc{tight} sample: ${}^{+0}_{-2\%}$;
({\em iii}) the statistics used to compute the \Etisol{}
distributions, and hence the precision of the matrix coefficients:
$\pm9\%$.  Effects ({\em i}) and ({\em ii}) are estimated as explained
in Section~\ref{sec:isolation-from-photons}.  Effect ({\em iii}) is
quantified by increasing and decreasing the $\epsilon,f$ parameters by
their statistical errors, and recomputing the signal yield: the
variations are then added in quadrature.

\subsubsection{Two-dimensional fit}
\label{sec:2DTemplate}

\begin{figure}
  \centering
  \includegraphics[width=\columnwidth]{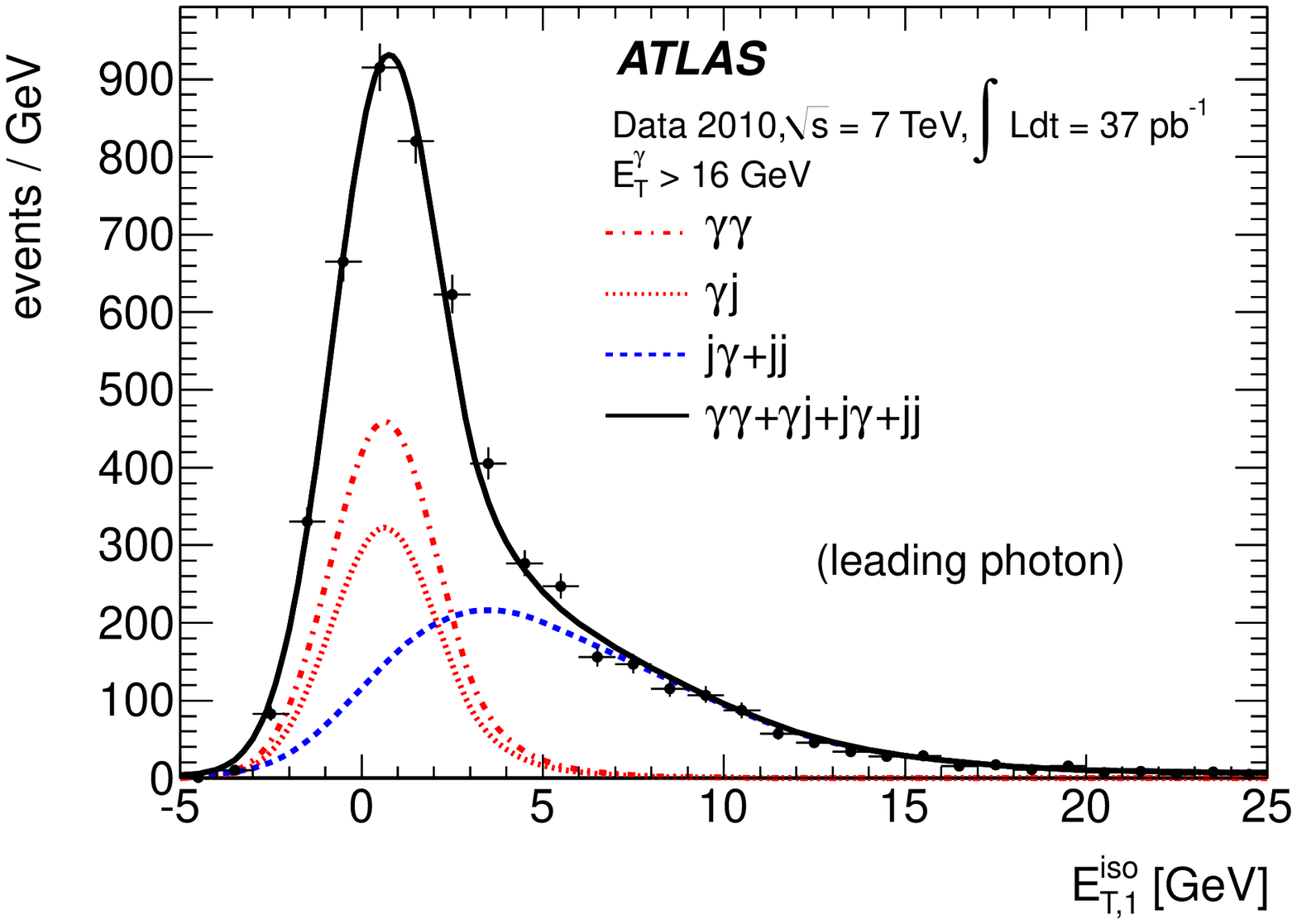}
  \includegraphics[width=\columnwidth]{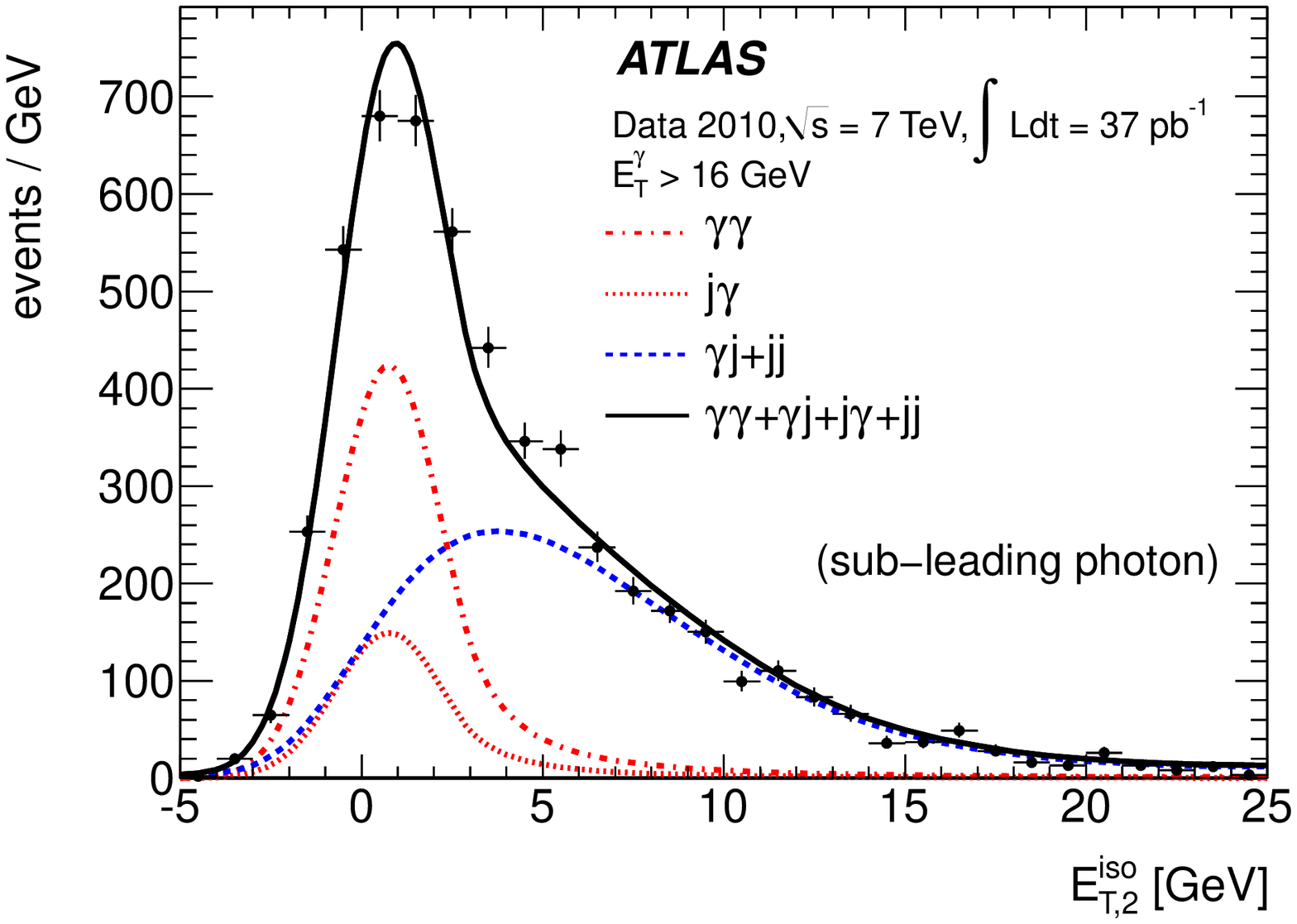}
  \caption{Projections of the 2-dimensional PDF fit on transverse
    isolation energies of the two photon candidates: leading photon
    (top) and sub-leading photon (bottom). Solid circles represent the
    observed data. The continuous curve is the fit result, while the
    dashed-dotted curve shows the \evtgamgam\ component.  The dashed
    line represents the background component of the leading and
    sub-leading photon sample, respectively }
  \label{fig:2Dfit_data}
\end{figure}

From all the di-photon events satisfying the \textsc{tight} selection
(sample \tight\tight), the observed 2-dimensional distribution
$F\obs(\EtisolOne,\EtisolTwo)$ of the isolation energies of the
leading and sub-leading photons is built.  Then, a linear combination
of four unbinned probability density functions (PDFs),
$F_\evtgamgam,F_\evtgamjet,F_\evtjetgam,F_\evtjetjet$, describing the
2-dimensional distributions of the four final states, is fit to the
observed distribution.  For the \evtgamgam, \evtgamjet, \evtjetgam\
final states, the correlation between $\EtisolOne$ and $\EtisolTwo$ is
assumed to be negligible, therefore the 2-dimensional PDFs are
factorized into the leading and sub-leading PDFs.  The leading and
sub-leading photon PDFs $F_{\objgam_1},F_{\objgam_2}$ are obtained
from the electron extrapolation, as described in
Section~\ref{sec:isolation-from-electrons}.  The background PDF
$F_{\objjet_2}$ for \evtgamjet{} events is obtained from the
non-\textsc{tight} sample on the sub-leading candidate, for events
where the leading candidate satisfies the \textsc{tight} selection.
The background PDF $F_{\objjet_1}$ for \evtjetgam{} events is obtained
in a similar way. Both background PDFs are then smoothed with
empirical parametric functions.  The PDF for \evtjetjet{} events
cannot be factorized, due to the sizable correlation between the two
candidates. Therefore, a 2-dimensional PDF is directly extracted from
events where both candidates belong to the non-\textsc{tight} sample,
then smoothed.

The four yields in the \tight\tight{} sample come from an extended maximum likelihood fit of:
\begin{eqnarray}
  N^{\tight\tight}F\obs(\EtisolOne,\EtisolTwo) 
  &=& N_\evtgamgam^{\tight\tight}F_{\objgam_1}(\EtisolOne)F_{\objgam_2}(\EtisolTwo) \nonumber \\
  &+& N_\evtgamjet^{\tight\tight}F_{\objgam_1}(\EtisolOne)F_{\objjet_2}(\EtisolTwo) \nonumber \\
  &+& N_\evtjetgam^{\tight\tight}F_{\objjet_1}(\EtisolOne)F_{\objgam_2}(\EtisolTwo) \nonumber \\
  &+& N_\evtjetjet^{\tight\tight}F_\evtjetjet(\EtisolOne,\EtisolTwo) \nonumber 
  ~.
\end{eqnarray}
Figure~\ref{fig:2Dfit_data} shows the fit result for the full \tight\tight{} data set.

The yields in the \tight\isol\tight\isol{} sample are evaluated by
multiplying $N_\evtgamgam^{\tight\tight}$ by the integral of the
2-dimensional signal PDF in the region defined by $\EtisolOne<3$~GeV
and $\EtisolTwo<3$~GeV.  The procedure is applied to the events
belonging to each bin of the observables \mgg, \ptgg, \dphigg.  The
result is displayed in Figure~\ref{fig:comparison}, by the open
triangles.

The main sources of systematic uncertainties are:
({\em i}) definition of the non-\textsc{tight} control sample: ${}^{+13\%}_{-0\%}$; 
({\em ii}) signal composition: $\pm8\%$;
({\em iii}) effect of material knowledge on signal: ${}^{+1.6\%}_{0\%}$;
({\em iv}) signal PDF parameters: $\pm0.7\%$;
({\em v}) jet PDF parameters: $\pm1.2\%$;
({\em vi}) di-jet PDF parameters: $\pm1\%$;
({\em vii}) signal contamination in the non-\textsc{tight} sample: ${}^{+1.2\%}_{0\%}$.
Effect ({\em i}) is estimated by changing the number of released strips cuts, as explained in Section~\ref{sec:isolation-from-photons}.
Effect ({\em ii}) has been estimated by artificially setting the fraction of fragmentation photons to 0\% or to 100\%.
Effect ({\em iii}) has been quantified by repeating the $\el\rightarrow\ph$ extrapolation based on Monte~Carlo samples with a distorted geometry.
Effects ({\em iv}, {\em v}) have been estimated by randomly varying the parameters of the smoothing functions, within their covariance 
ellipsoid, and repeating the 2-dimensional fit. Effect ({\em vi}) has been estimated by randomly extracting a set of $(\EtisolOne,\EtisolTwo)$ 
pairs, comparable to the experimental statistics, from the smoothed 
$F_\evtjetjet$ PDF, then re-smoothing the obtained distribution and repeating the 2-dimensional fit. 
Effect ({\em vii}) has been estimated by taking the signal contamination from simulation --- neglected when computing the central value.

\subsubsection{Isolation vs identification sideband counting (2D-sidebands)}
\label{sec:2DSidebands}

\begin{figure}
  \centering
  \includegraphics[width=0.9\columnwidth]{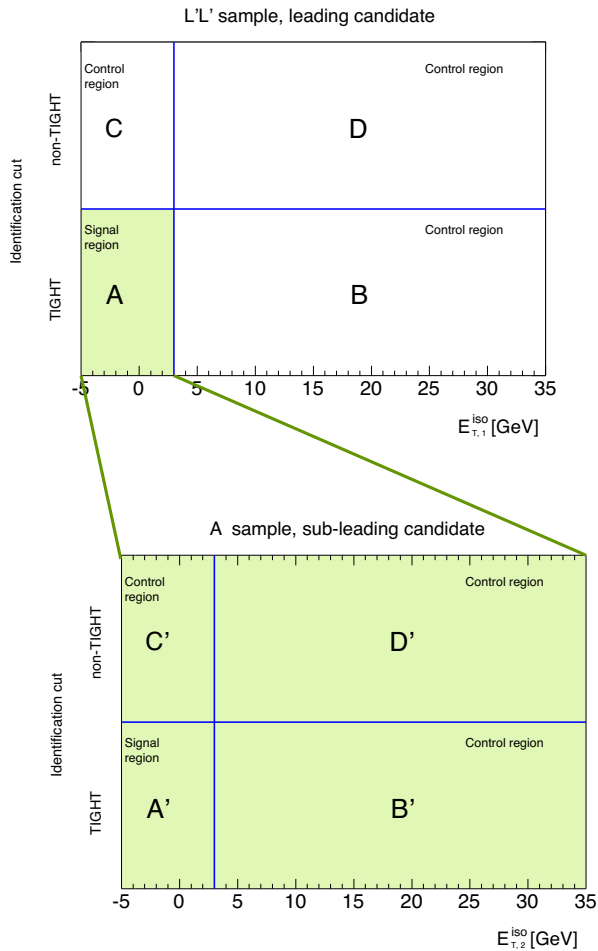}
  \caption{Schematic representation of the two-dimensional sideband
    method.  The top plane displays the isolation ($x$-axis) and
    \textsc{tight} identification ($y$-axis) criteria for the
    classification of the leading photon candidate.  When the leading
    photon belongs to region $A$, the same classification is applied
    to the sub-leading photon, as described by the bottom plane.  }
  \label{fig:2D-S_2}
\end{figure}

This method has been used in ATLAS in the inclusive photon
cross-section measurement~\cite{ATLAS-INCLPHOTON} and in the
background decomposition in the search for the Higgs boson decaying
into two photons~\cite{ATLAS_Higgs_all}.

The base di-photon sample must fulfil the selection with the strips
cuts released, defined by the union of \textsc{tight} and
non-\textsc{tight} samples and here referred to as \textsc{loose'}
(\loose').  The leading photons in the \loose'\loose' sample are
divided into four categories $A,~B,~C,~D$, depending on whether they
satisfy the \textsc{tight} selection and/or the isolation requirement
--- see Figure~\ref{fig:2D-S_2}~(top).  The signal region, defined by
\textsc{tight} and isolated photons (\tight\isol), contains $\NA$
candidates, whereas the three control regions contain $\NB,~\MA,~\MB$
candidates.  Under the hypothesis that regions $B,~C,~D$ are largely
dominated by background, and that the isolation energy of the
background has little dependence on the \textsc{tight} selection (as
discussed in Section~\ref{sec:isolation-from-photons}), the number of
genuine leading photons $\NAsig$ in region $A$, coming from
\evtgamgam\ and \evtgamjet\ final states, can be
computed~\cite{ATLAS-INCLPHOTON} by solving the equation:
\begin{equation}
  \NAsig = \NA - \left[ (\NB-c_1\NAsig) \frac{\MA-c_2\NAsig}{\MB-c_1c_2\NAsig}\right]R\bkg \label{eq:2DS-NAsig} 
  ~.
\end{equation}
Here, $c_1$ and $c_2$ are the signal fractions failing respectively
the isolation requirement and the \textsc{tight} selection. The former
is computed from the isolation distributions, as extracted in
Section~\ref{sec:isolation-from-photons}; the latter is evaluated from
Monte~Carlo simulation, after applying the corrections to adapt it to
the experimental shower shapes
distributions~\cite{ATLAS-INCLPHOTON}. The parameter
$R\bkg=\frac{\NAbkg\MBbkg}{\MAbkg\NBbkg}$ measures the degree of
correlation between the isolation energy and the photon selection in
the background: it is set to $1$ to compute the central values, then
varied according to the ``di-jet-like'' Monte~Carlo prediction for
systematic studies.

When the leading candidate is in the \tight\isol{} region, the
sub-leading one is tested, and four categories $A',~B',~C',~D'$ are
defined, as in the case of the leading candidate --- see
Figure~\ref{fig:2D-S_2}~(bottom). The number of genuine sub-leading
photons $\NAPsig$, due to \evtgamgam\ and \evtjetgam\ final states, is
computed by solving an equation analogous to~(\ref{eq:2DS-NAsig}).

$\NAsig$ and $\NAPsig$ are related to the yields by:
\begin{eqnarray*}
  \NAsig &=& \frac{N_\evtgamgam^{\tight\isol\tight\isol}}{\epsilon'} + \frac{N_\evtgamjet^{\tight\isol\tight\isol}}{f'} ~,\\
  \NAPsig &=& N_\evtgamgam^{\tight\isol\tight\isol} + N_\evtjetgam^{\tight\isol\tight\isol} ~,
\end{eqnarray*}
where 
$\epsilon'=\frac{1}{(1+c'_1)(1+c'_2)}$ 
is the probability that a sub-leading photon satisfies the \textsc{tight} selection and isolation requirement,
while $f'$ is the analogous probability for a jet faking a sub-leading photon. 
The di-photon yield is therefore computed as:
\begin{equation}
  N_\evtgamgam^{\tight\isol\tight\isol} = \frac{\epsilon' \left(\alpha f' \NAsig + (\alpha-1)\NAPsig\right)}
  {(\alpha-1) \epsilon' + \alpha f'} ~,
  \label{eq:2DS-yield}
\end{equation}
and $f'$ can be computed from the observed quantities to be $ f'=
\frac{\NAP - \NAPsig}{\NA - \NAPsig/\epsilon'} $.  The parameter
$\alpha$ is defined as the fraction of photon-jet events in which the
jet fakes the leading photon,
$\alpha=\frac{N_\evtjetgam^{\tight\isol\tight\isol}}{N_\evtgamjet^{\tight\isol\tight\isol}+N_\evtjetgam^{\tight\isol\tight\isol}}$,
whose value is taken from the \textsc{Pythia} photon-jet simulation.

The counts $\NA$, $\NB$, $\MA$, $\MB$, $\NAP$, $\NBP$, $\MAP$, $\MBP$,
and hence the yield, can be computed for all events entering a given
bin of \mgg, \ptgg, \dphigg.  The result is displayed in
Figure~\ref{fig:comparison}, by the open squares.

The main source of systematic error is the definition of the
non-\textsc{tight} sample: it induces an error of
${}^{+7\%}_{-10\%}$. The other effects come from the uncertainties of
the parameters entering equation~(\ref{eq:2DS-yield}). The main
effects come from: ({\em i}) variation of $c'_1$: $\pm4\%$; ({\em ii})
variation of $\alpha$: $\pm3\%$; ({\em iii}) variations of
$R\bkg,~{R'}\bkg$: ${}^{+0\%}_{-1.5\%}$.  The variations of
$c_1,~c_2,~c'_2$ have negligible impact.

\subsection{Electron background}
\label{sec:ElectronBkd}

\begin{figure}
  \centering
  \includegraphics[width=\columnwidth]{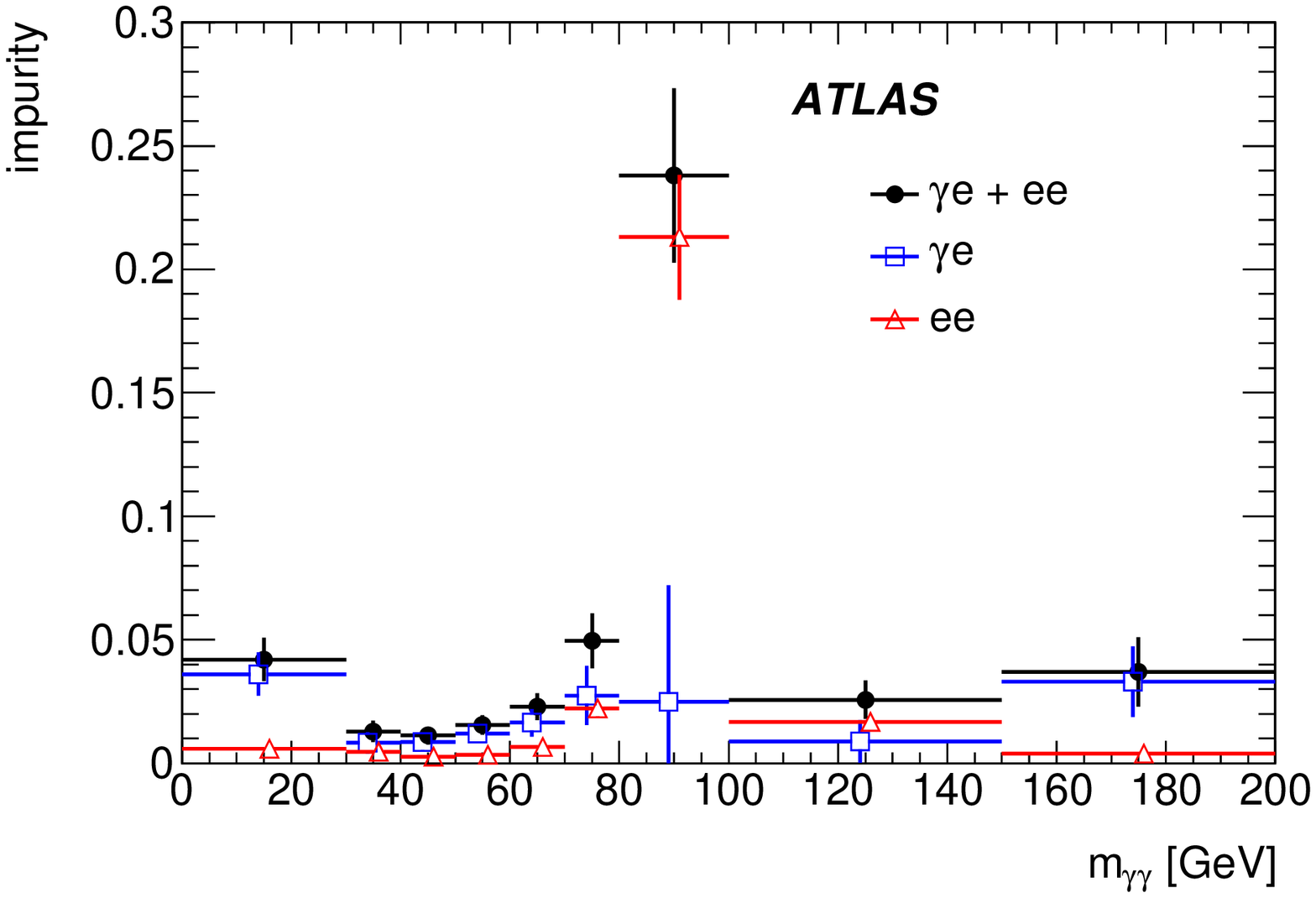}
  \includegraphics[width=\columnwidth]{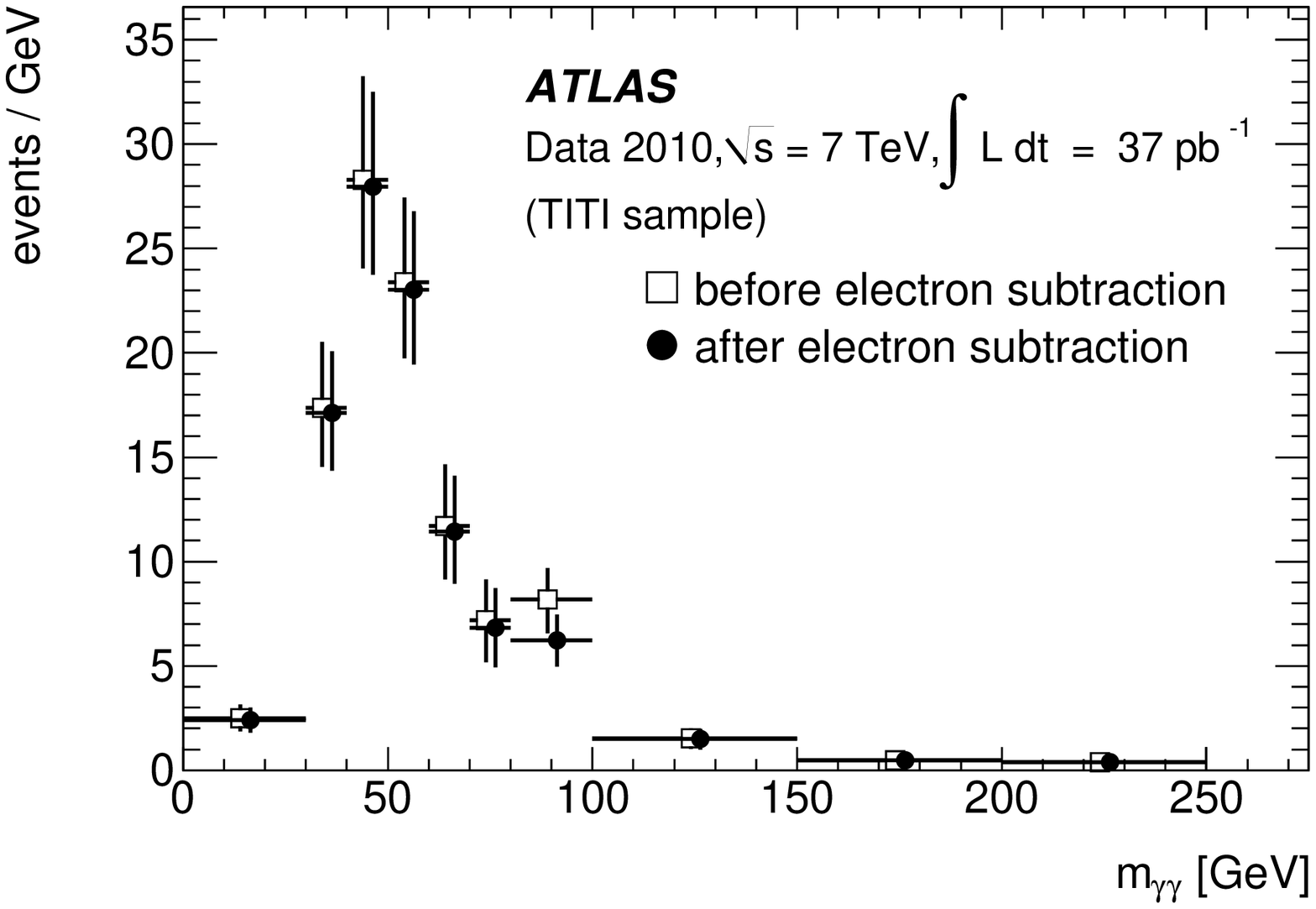}
  \caption{Electron background subtraction as a function of $m_{\ph\ph}$. The top plot displays the impurity,
    overall and for the $\ph\el$ and $\el\el$ separately. The bottom plot shows the di-photon yield before (open squares) and
    after (solid circles) the electron background subtraction.
    The points are artificially shifted horizontally, to better display the different values.
  }
  \label{fig:imp_mass}
\end{figure}

Background from isolated electrons contaminates mostly the selected
converted photon sample.  The contamination in the di-photon analysis
comes from several physical channels: ({\em i}) $e^+e^-$ final states
from Drell-Yan processes, $Z\rightarrow e^+e^-$ decay,
$W^+W^-\rightarrow e^+e^- \nu\bar{\nu}$; ({\em ii}) $\gamma e^\pm$
final states from di-boson production, e.g. $\gamma
W^\pm\rightarrow\gamma e^\pm\nu$, $\gamma Z\rightarrow\gamma e^+e^-$.
The effect of the $Z\rightarrow e^+e^-$ contamination is visible in
Figure~\ref{fig:comparison} in the mass bin $80<\mgg<100$~GeV.

Rather than quantifying each physical process separately, a global
approach is chosen. The events reconstructed with $\ph\ph$, $\ph\el$
and $\el\el$ final states are counted, thus obtaining counts
$N_{\ph\ph}$, $N_{\ph\el}$ and $N_{\el\el}$. Only photons and
electrons satisfying a \textsc{tight} selection and the calorimetric
isolation $\Etisol<3$~GeV are considered, and electrons are counted
only if they are not reconstructed at the same time as photons.  Such
counts are related to the actual underlying yields
$N\true_{\ph\ph},N\true_{\ph\el},N\true_{\el\el}$, defined as the
number of reconstructed final states where both particles are
correctly classified. Introducing the ratio
$\fph=\frac{N_{\el\rightarrow\ph}}{N_{\el\rightarrow\el}}$ between
genuine electrons that are wrongly and correctly classified, and
likewise $\fel=\frac{N_{\ph\rightarrow\el}}{N_{\ph\rightarrow\ph}}$
for genuine photons, the relationship between the $N$ and $N\true$
quantities is described by the following linear system:
\begin{equation}
  {\scriptsize
    \left(\begin{array}{c} N_{\ph\ph} \\ N_{\ph\el} \\ N_{\el\el} \end{array}\right) =
    \left(\begin{array}{ccc} 
        1        & \fph           & (\fph)^2 \\
        2\fel    & (1+\fph\fel)   & 2\fph    \\
        (\fel)^2 & \fel           & 1        \\
      \end{array}\right) \times 
    \left(\begin{array}{c} N\true_{\ph\ph} \\ N\true_{\ph\el} \\ N\true_{\el\el} \end{array}\right)
    \label{eq:elbkg-N-from-Ntrue}
  }
\end{equation}
which can be solved for the unknown $N\true_{\ph\ph}$.

The value of \fph\ is extracted from collision data, as
$\fph=\frac{N_{\ph\el}}{2N_{\el\el}}$, from events with an invariant
mass within $\pm5$~GeV of the $Z$ mass. The continuum background is
removed using symmetric sidebands. The result is
$\fph=0.112\pm0.005(\textrm{stat})\pm0.003(\textrm{syst})$, where the
systematic error comes from variations of the mass window and of the
sidebands.  This method has been tested on ``di-jet-like'' and
$Z\rightarrow e^+e^-$ Monte~Carlo samples and shown to be unbiased.
The value of \fel\ is taken from the ``di-jet-like'' Monte~Carlo:
$\fel=0.0077$.  To account for imperfect modelling, this value has
also been set to 0, or to three times the nominal value, and the
resulting variations are considered as a source of systematic error.

The electron contamination is estimated for each bin of \mgg, \ptgg\
and \dphigg, and subtracted from the di-photon yield.  The result, as
a function of \mgg, is shown in Figure~\ref{fig:imp_mass}. The
fractional contamination as a function of \ptgg\ and \dphigg\ is
rather flat, amounting to $\sim5\%$.

\section{Efficiencies and unfolding}
\label{sec:efficiency}

The signal is defined as a di-photon final state, which must satisfy
precise kinematic cuts (referred to as ``fiducial acceptance''):
\Itemize{
\item both photons must have a transverse momentum $\pt^\gamma>16$~GeV
  and must be in the pseudorapidity acceptance $|\eta^\gamma|<2.37$,
  with the exclusion of the region $1.37<|\eta^\gamma|<1.52$;
\item the separation between the two photons must be $\Delta
  R_\evtgamgam=\sqrt{(\eta_1^\gamma-\eta_2^\gamma)^2+(\phi_1^\gamma-\phi_2^\gamma)^2}>0.4$;
\item both photons must be isolated, i.e.\ the transverse energy flow
  \Etpartisol\ due to interacting particles in a cone of angular
  radius $R<0.4$ must be $\Etpartisol<4$~GeV.  }
These kinematic cuts define a phase space similar to the experimental
selection described in Section~\ref{sec:Selection}.  In particular,
the requirement on \Etpartisol\ has been introduced to match
approximately the experimental cut on \Etisol.  The value of
\Etpartisol\ is corrected for the ambient energy, similarly to what is
done for \Etisol. From studies on a \textsc{Pythia} di-photon
Monte~Carlo sample, there is a high correlation between the two
variables, and $\Etisol=3$~GeV corresponds to
$\Etpartisol\simeq4$~GeV.

A significant number of di-photon events lying outside the fiducial
acceptance pass the experimental selection because of resolution
effects: these are referred to as ``below threshold'' ($BT$) events.

The background subtraction provides the di-photon signal yields for
events passing all selections (\tight\isol\tight\isol).  Such yields
are called $N^{\tight\isol\tight\isol}_i$, where the index $i$ flags
the bins of the reconstructed observable $X\rec$ under consideration
($X$ being \mgg, \ptgg, \dphigg).  The relationship between
$N^{\tight\isol\tight\isol}_i$ and the true yields $n_\alpha$
($\alpha$ being the bin index of the true value $X\true$) is:
\begin{eqnarray}
  N^{\tight\isol\tight\isol}_i &=& \epsilon^{trigger} \epsilon^{\tight\tight}_i N^{\isol\isol}_i ~,\\
  N^{\isol\isol}_i\left(1-f^{BT}_i\right) &=& \sum_{\alpha} M_{i\alpha}\effRA_\alpha n_\alpha ~,
\end{eqnarray}
where $N^{\isol\isol}_i$ is the number of reconstructed isolated
di-photon events in the $i$-th bin, and: 
\Itemize{
\item $\epsilon^{trigger}$ is the trigger efficiency, computed for
  events where both photons satisfy the \textsc{tight} identification
  and the calorimetric isolation;
\item $\epsilon^{\tight\tight}_i$ is the efficiency of the
  \textsc{tight} identification, for events where both photons satisfy
  the calorimetric isolation;
\item $f^{BT}_i$ is the fraction of ``below-threshold'' events;
\item $M_{i\alpha}$ is a ``migration probability'', i.e. the
  probability that an event with $X\true$ in bin-$\alpha$ is
  reconstructed with $X\rec$ in bin-$i$;
\item $\effRA_\alpha$ accounts for both the reconstruction efficiency
  and the acceptance of the experimental cuts (kinematics and calorimetric isolation).
}

\subsection{Trigger efficiency}
\label{sec:eff-trigger}

The trigger efficiency is computed from collision data, for events
containing two reconstructed photons with transverse energy
$\et^\gamma>16$~GeV, both satisfying the \textsc{tight} identification
and the calorimetric isolation requirement
(\tight\isol\tight\isol). The computation is done in three steps.

First, a level-1 $e/\gamma$ trigger with an energy threshold of 5~GeV
is studied: its efficiency, for reconstructed \tight\isol\ photons, is
measured on an inclusive set of minimum-bias events: for
$\et^\gamma>16$~GeV it is $\epsilon_0=100.0^{+0.0}_{-0.1}\%$ ---
therefore such a trigger does not bias the sample.  Next, a high-level
photon trigger with a 15~GeV threshold is studied, for reconstructed
\tight\isol\ photons selected by the level-1 trigger: its efficiency
is $\epsilon_1=99.1^{+0.3}_{-0.4}\%$ for $\et^\gamma>16$~GeV. Finally,
di-photon \tight\isol\tight\isol\ events with the sub-leading photon
selected by a high-level photon trigger are used to compute the
efficiency of the di-photon 15~GeV-threshold high-level trigger,
obtaining $\epsilon_2=99.4^{+0.5}_{-1.0}\%$.  The overall efficiency
of the trigger is therefore
$\epsilon^{trigger}=\epsilon_0\epsilon_1\epsilon_2=(98.5^{+0.6}_{-1.0}\pm
1.0)\%$.  The first uncertainty is statistical, the second is
systematic and accounts for the contamination of photon-jet and di-jet
events in the selected sample.

\subsection{Identification efficiency}
\label{sec:eff-ID}

The photon \textsc{tight} identification efficiency
$\epsilon^{\tight|\isol}$, for photon candidates satisfying the
isolation cut $\Etisol<3$~GeV, is computed as described in
Ref~\cite{ATLAS-INCLPHOTON}, as a function of $\eta^\gamma$ and
$\et^\gamma$. The efficiency is determined by applying the
\textsc{tight} selection to a Monte~Carlo photon sample, where the
shower shape variables have been shifted to better reproduce the
observed distributions. The shift factors are obtained by comparing
the shower shapes of photon candidates from a ``di-jet-like''
Monte~Carlo sample to those observed in collision data.  To enhance
the photon component in the sample --- otherwise overwhelmed by the
jet background --- only the photon candidates satisfying the
\textsc{tight} selection are considered. This procedure does not bias
the bulk of the distribution under test appreciably, since the cuts
have been tuned to reject only the tails of the photons'
distributions.  However, to check the systematic effect due to the
selection, the shift factors are also recomputed applying the
\textsc{loose} selection.

Compared to Ref~\cite{ATLAS-INCLPHOTON},
the photon identification cuts have been re-optimized to reduce the systematic errors, and converted and unconverted photons
treated separately. The photon identification efficiency is $\eta^\gamma$-dependent, and increases with $\et^\gamma$, ranging from $\sim60\%$ for $16<\et^\gamma<20$~GeV,
to $\gtrsim90\%$ for $\et^\gamma>100$~GeV. The overall systematic error is between 2\% and 10\%, the higher values being applicable at 
lower $\et^\gamma$ and for converted photons. The main sources of systematic uncertainty are: 
({\em i}) the systematic error on the shift factors; 
({\em ii}) the knowledge of the detector material;
({\em iii}) the failure to detect a conversion, therefore applying the wrong \textsc{tight} identification.

Rather than computing an event-level identification efficiency for
each bin of each observable, the photon efficiency can be naturally
accommodated into the event weights described in
Section~\ref{sec:4x4Matrix}, by dividing the weight $w^{(k)}$ of
equation~(\ref{eq:4x4-weight-TITI}) by the product of the two photon
efficiencies:
\begin{equation}
  N_i^{\isol\isol} = \sum_{k\in\textrm{bin-$i$}} \frac{w^{(k)}}
  {\left[\epsilon^{\tight|\isol}(\eta_1^\gamma,{\et}_1^\gamma)\epsilon^{\tight|\isol}(\eta_2^\gamma,{\et}_2^\gamma)\right]^{(k)}}
  ~,
  \label{eq:4x4-yield-II}
\end{equation}
where the sum is extended over all events in the \tight\tight\ sample
and in the $i$-th bin.  Here the identification efficiencies of the
two photons are assumed to be uncorrelated --- which is ensured by the
separation cut $\Delta R>0.4$, and by the binning in $\eta^\gamma$ and
$\et^\gamma$.

The event efficiency,
$\epsilon^{\tight\tight}_i=\frac{N_i^{\tight\isol\tight\isol}}{N_i^{\isol\isol}}$,
is essentially flat at $\sim60\%$ in \dphigg, and increases with \mgg\
and \ptgg, ranging from $\sim55\%$ to $\sim80\%$. Its total systematic
error is $\sim10\%$, rather uniform over the \mgg, \ptgg, \dphigg\
ranges.

\subsection{Reconstruction, acceptance, isolation and unfolding}
\label{sec:eff-rec-and-unfolding}

The efficiency $\effRA_\alpha$ accounts for both the reconstruction
efficiency and the acceptance of the experimental selection. It is
computed for each bin of $X\true$, with Monte~Carlo di-photon events
generated with \textsc{Pythia} in the fiducial acceptance, as the
fraction of events where both photons are reconstructed, pass the
acceptance cuts and the calorimetric isolation.  The value of
$\effRA_\alpha$ ranges between 50\% and 60\%. The two main sources of
inefficiency are the local ECAL read-out failures ($\sim -18\%$) and
the calorimetric isolation ($\sim -20\%$).

The energy scale differences between Monte~Carlo and collision data
--- calibrated on $Z\rightarrow e^+e^-$ events --- are taken into
account.  The uncertainties on the energy scale and resolution are
propagated as systematic errors through the evaluation: the former
gives an effect between $+3\%$ and $-1\%$ on the signal rate, while
the latter has negligible impact.

In Monte~Carlo, the calorimetric isolation energy, \Etisol, needs to
be corrected to match that observed in collision data.  The correction
is optimized on \textsc{tight} photons, for which the background
contamination can be removed (see
Section~\ref{sec:isolation-from-photons}), then it is applied to all
photons in the Monte~Carlo sample.  The \Etisol\ difference observed
between Monte~Carlo simulation and collision data may be entirely due
to inaccurate \textsc{Geant4}/detector modelling, or it can also be a
consequence of the physical model in the generator (e.g. kinematics,
fragmentation, hadronization).  From the comparison between collision
data and simulation, the two effects cannot be disentangled.  To
compute the central values of the results, the difference between
simulation and collision data is assumed to be entirely due to the
detector simulation.  As a cross-check, the opposite case is assumed:
that the difference is entirely due to the generator model. In this
case, the particle-level isolation \Etpartisol\ should also be
corrected, using the $\Etpartisol\rightarrow\Etisol$ relationship
described by the detector simulation. This modifies the definition of
fiducial acceptance, and hence the values of $\effRA_\alpha$,
resulting in a cross-section variation of $\sim-7\%$, which is handled
as an asymmetric systematic uncertainty.

The fraction of events ``below threshold'', $f^{BT}_i$, is computed
from the same \textsc{Pythia} signal Monte~Carlo sample, for each bin
of $X\rec$.  Its value is maximum ($\sim12\%$) for \mgg\ about twice
the \et\ cut, and decreases to values $<5\%$ for $\mgg>50$~GeV.

The ``migration matrix'', $M_{i\alpha}$, is filled with
\textsc{Pythia} Monte~Carlo di-photon events in the fiducial
acceptance, that are reconstructed, pass the acceptance cuts and the
calorimetric isolation. The inversion of this matrix is performed with
an unfolding technique, based on Bayesian
iterations~\cite{iterative-unfolding}.  The systematic uncertainties
of the procedure have been estimated with a large number of toy
datasets and found to be negligible. The result has also been tested
to be independent of the initial (``prior'') distributions. Moreover,
it has been checked that a simpler bin-by-bin unfolding yields
compatible results.

As the evaluation of $\effRA_\alpha,~f^{BT}_i,~M_{i\alpha}$ may
strongly depend on the simulation modelling, two additional Monte
Carlo samples have been used, the first with more material modelled in
front of the calorimeter, and the second with a different generator
(\textsc{Sherpa}): the differences on the computed signal rates are
$\sim+10\%$ and $\lesssim+5\%$ respectively, and are treated as
systematic errors.

\section{Cross-section measurement}
\label{sec:cross-section}

The di-photon production cross-section is evaluated from the corrected
binned yields $n_\alpha$, divided by the integrated luminosity $\int
L\de t=(37.2\pm1.3)~\ipb$~\cite{ATLAS_LUMI_all}.  The results are
presented as differential cross-sections, as functions of the three
observables \mgg, \ptgg, \dphigg, for a phase space defined by the
fiducial acceptance cuts in Section~\ref{sec:efficiency}.  In
Table~\ref{tab:cross-sections-tot}, the differential cross-section is
quoted for each bin, with its statistical and systematic
uncertainty. In Table~\ref{tab:cross-sections-breakdown}, all the
considered sources of systematic errors are listed separately.

\begin{figure}
  \begin{center}
    \includegraphics[width=\columnwidth]{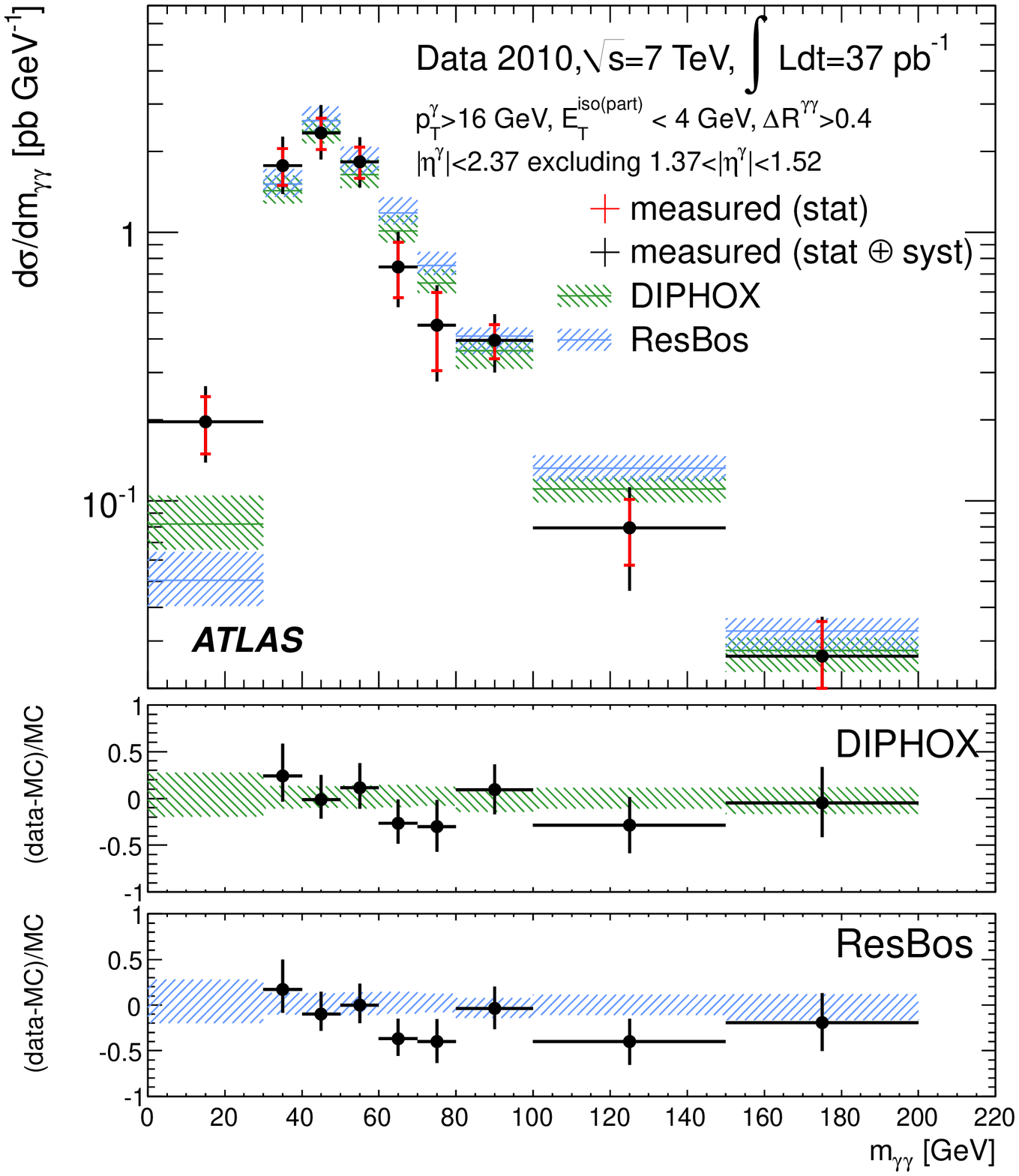}
    \caption{
      Differential cross-section $\de\sigma/\de\mgg$ of di-photon production. 
      The solid circles display the experimental values,
      the hatched bands display the NLO computations by DIPHOX and ResBos.
      The bottom panels show the relative difference between the measurements and the NLO predictions.
      The data/theory point in the bin $0<\mgg<30$~GeV lies above the frames.
    }
    \label{fig:cross-section-mass}
  \end{center}
\end{figure}

\begin{figure}
  \begin{center}
    \includegraphics[width=\columnwidth]{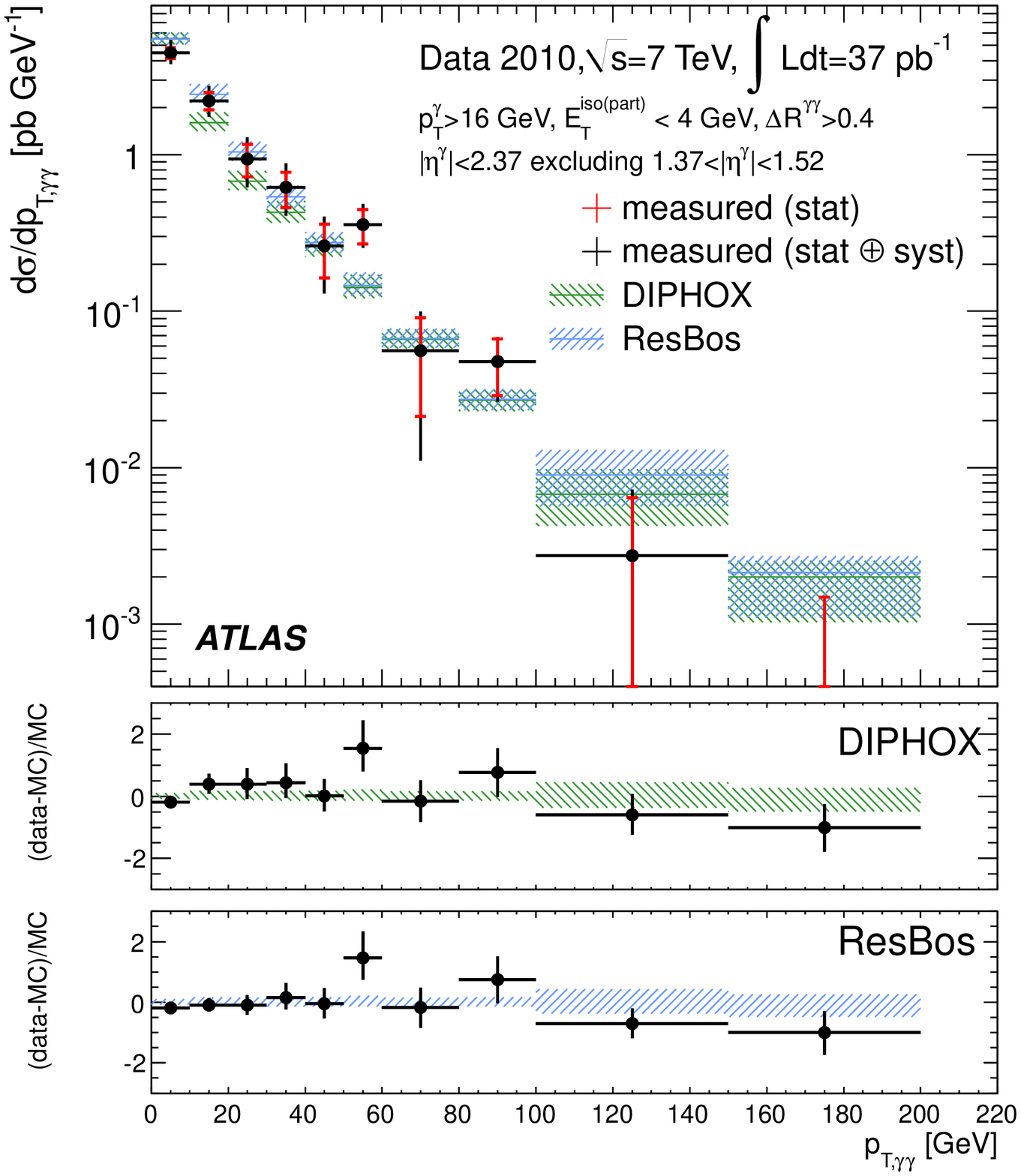}
    \caption{
      Differential cross-section $\de\sigma/\de\ptgg$ of di-photon production. 
      The solid circles display the experimental values,
      the hatched bands display the NLO computations by DIPHOX and ResBos.
      The bottom panels show the relative difference between the measurements and the NLO predictions.
      The data point in the bin $150<\ptgg<200$~GeV in the main panel lies below the frame.
    }
    \label{fig:cross-section-ptgg}
  \end{center}
\end{figure}

\begin{figure}[t]
  \begin{center}
    \includegraphics[width=\columnwidth]{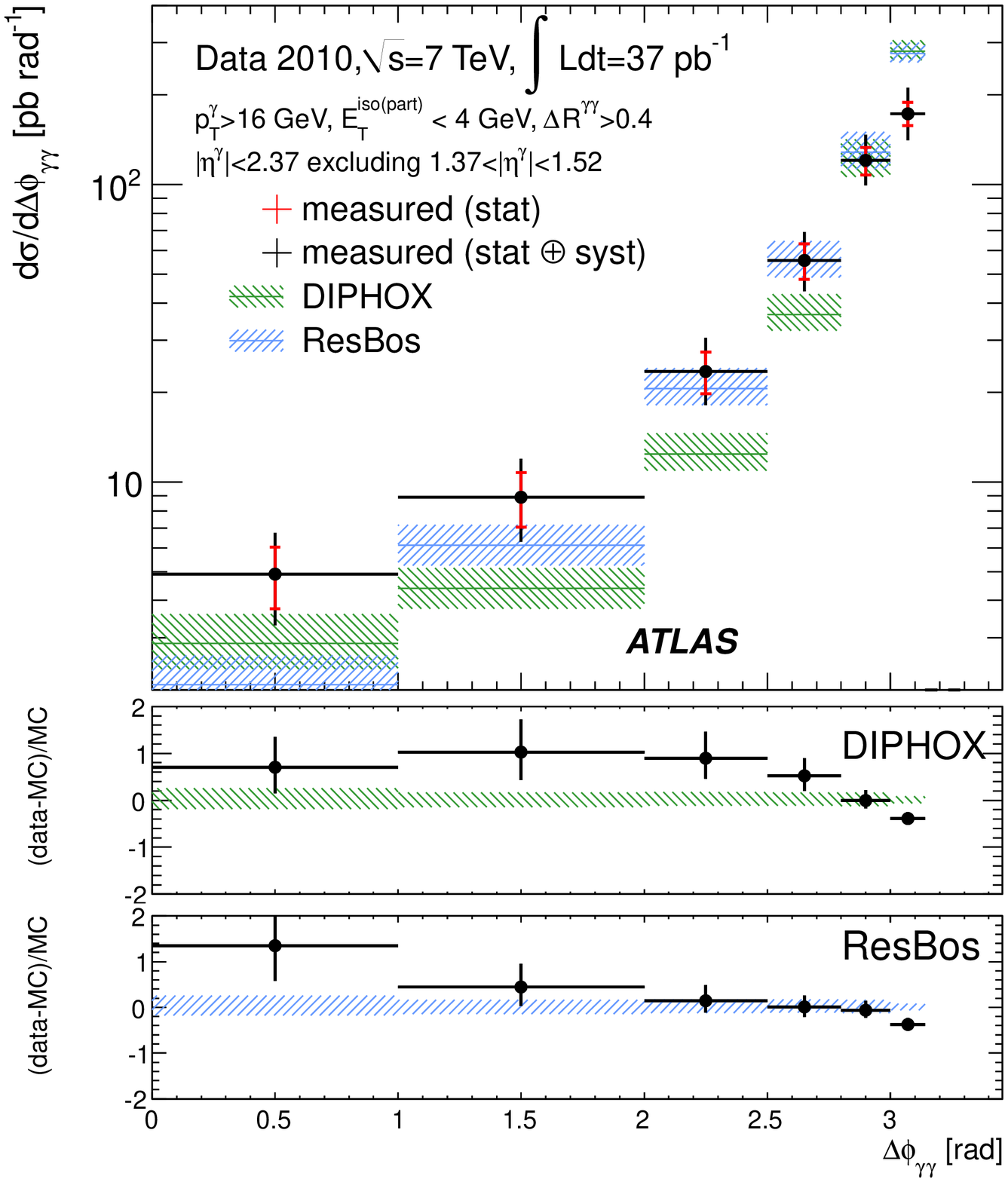}
    \caption{
      Differential cross-section $\de\sigma/\de\dphigg$ of di-photon production. 
      The solid circles display the experimental values,
      the hatched bands display the NLO computations by DIPHOX and ResBos.
      The bottom panels show the relative difference between the measurements and the NLO predictions.
    }
    \label{fig:cross-section-dphi}
  \end{center}
\end{figure}

The experimental measurement is compared with theoretical predictions
from the DIPHOX~\cite{diphox} and \mbox{ResBos}~\cite{Balazs:2007hr}
NLO generators in Figures~\ref{fig:cross-section-mass},
\ref{fig:cross-section-ptgg} and~\ref{fig:cross-section-dphi}.  The
DIPHOX and ResBos evaluation has been carried out using the NLO
fragmentation function~\cite{Bourhis:2000gs} and the CTEQ6.6 parton
density function (PDF) set~\cite{cteq}.  The fragmentation,
normalization and factorization scales are set equal to \mgg. The same
fiducial acceptance cuts introduced in the signal definition
(Section~\ref{sec:efficiency}) are applied. Since neither generator
models the hadronization, it is not possible to apply a requirement on
\Etpartisol: the closest isolation variable available in such
generators is the ``partonic isolation'', therefore this is required
to be less then 4~\GeV. The computed cross-section shows a weak
dependence on the partonic isolation cut: moving it to 2~GeV or 6~GeV
produces variations within 5\%, smaller than the theoretical
systematic errors.

The theory uncertainty error bands come from scale and PDF
uncertainties evaluated from DIPHOX: ({\em i}) variation of
renormalization, fragmentation and factorization scales: each is
varied to $\frac{1}{2}\mgg$ and $2\mgg$, and the envelope of all
variations is assumed as a systematic error; ({\em ii}) variation of
the eigenvalues of the PDFs: each is varied by $\pm1\sigma$, and
positive/negative variations are summed in quadrature separately.  As
an alternative, the MSTW~2008 PDF set has been used: the difference
with respect to CTEQ6.6 is an overall increase by $\sim10\%$, which is
covered by the CTEQ6.6 total systematic error.

The measured distribution of $\de\sigma/\de\dphigg$
(Figure~\ref{fig:cross-section-dphi}) is clearly broader than the
DIPHOX and ResBos predictions: more photon pairs are seen in data at
low \dphigg\ values, while the theoretical predictions favour a larger
back-to-back production ($\dphigg\simeq\pi$).  This result is
qualitatively in agreement with previous measurements at the
Tevatron~\cite{D0_diphoton,CDF_diphoton_2011}.  The distribution of
$\de\sigma/\de\mgg$ (Figure~\ref{fig:cross-section-mass}) 
agrees within the assigned uncertainties with both the
DIPHOX and ResBos predictions, apart from the
region $\mgg<2\et^\mathrm{cut}$ ($\et^\mathrm{cut}=16$~GeV being the
applied cut on the photon transverse momenta): as this region is
populated by events with small \dphigg, the poor quality of the
predictions can be related to the discrepancy observed in the \dphigg\
distribution.  The result for $\de\sigma/\de\ptgg$
(Figure~\ref{fig:cross-section-ptgg}) 
is in agreement with both DIPHOX
and ResBos: the maximum deviation, about $2\sigma$, is observed in the
region $50<\ptgg<60$~GeV.

\begin{table}
  \begin{center}
    \caption{Binned differential cross-sections $\de\sigma/\de\mgg$,
      $\de\sigma/\de\ptgg$, $\de\sigma/\de\dphigg$ for di-photon production.
      For each bin, the differential cross-section is quoted with its statistical
      and systematic uncertainties (symmetric and asymmetric, respectively).
      Values quoted as $0.000$ are actually less than $0.0005$ in absolute value.
    }
    \label{tab:cross-sections-tot}
    \setlength{\extrarowheight}{4pt}
\begin{tabular}{crll}
\hline\hline
$\mgg~\mathrm{[GeV]}$ & \multicolumn{3}{c}{$d\sigma/d\mgg~\mathrm{[pb/GeV]}$} \\
[4pt]\hline
$     0~-~    30$~~~  & $0.20$ & $\pm~$0.05 & ${}^{+0.05}_{-0.03}$  \\
$    30~-~    40$~~~  & $1.8$ & $\pm~$0.3 & ${}^{+0.4}_{-0.3}$  \\
$    40~-~    50$~~~  & $2.3$ & $\pm~$0.3 & ${}^{+0.6}_{-0.4}$  \\
$    50~-~    60$~~~  & $1.83$ & $\pm~$0.24 & ${}^{+0.36}_{-0.28}$  \\
$    60~-~    70$~~~  & $0.74$ & $\pm~$0.17 & ${}^{+0.19}_{-0.13}$  \\
$    70~-~    80$~~~  & $0.45$ & $\pm~$0.15 & ${}^{+0.11}_{-0.09}$  \\
$    80~-~   100$~~~  & $0.40$ & $\pm~$0.06 & ${}^{+0.08}_{-0.08}$  \\
$   100~-~   150$~~~  & $0.079$ & $\pm~$0.022 & ${}^{+0.025}_{-0.025}$  \\
$   150~-~   200$~~~  & $0.026$ & $\pm~$0.009 & ${}^{+0.006}_{-0.004}$  \\
[4pt]\hline\hline
$\ptgg~\mathrm{[GeV]}$ & \multicolumn{3}{c}{$d\sigma/d\ptgg~\mathrm{[pb/GeV]}$} \\
[4pt]\hline
$     0~-~    10$~~~  & $4.5$ & $\pm~$0.4 & ${}^{+0.9}_{-0.6}$  \\
$    10~-~    20$~~~  & $2.2$ & $\pm~$0.3 & ${}^{+0.5}_{-0.4}$  \\
$    20~-~    30$~~~  & $0.94$ & $\pm~$0.22 & ${}^{+0.28}_{-0.24}$  \\
$    30~-~    40$~~~  & $0.62$ & $\pm~$0.16 & ${}^{+0.21}_{-0.14}$  \\
$    40~-~    50$~~~  & $0.26$ & $\pm~$0.10 & ${}^{+0.10}_{-0.09}$  \\
$    50~-~    60$~~~  & $0.36$ & $\pm~$0.09 & ${}^{+0.09}_{-0.05}$  \\
$    60~-~    80$~~~  & $0.06$ & $\pm~$0.03 & ${}^{+0.03}_{-0.03}$  \\
$    80~-~   100$~~~  & $0.048$ & $\pm~$0.019 & ${}^{+0.009}_{-0.010}$  \\
$   100~-~   150$~~~  & $0.003$ & $\pm~$0.004 & ${}^{+0.003}_{-0.002}$  \\
$   150~-~   200$~~~  & $0.000$ & $\pm~$0.002 & ${}^{+0.000}_{-0.000}$  \\
[4pt]\hline\hline
$\dphigg~\mathrm{[rad]}$ & \multicolumn{3}{c}{$d\sigma/d\dphigg~\mathrm{[pb/rad]}$} \\
[4pt]\hline
$  0.00~-~  1.00$~~~  & $4.9$ & $\pm~$1.1 & ${}^{+1.5}_{-1.1}$  \\
$  1.00~-~  2.00$~~~  & $8.9$ & $\pm~$1.8 & ${}^{+2.5}_{-1.9}$  \\
$  2.00~-~  2.50$~~~  & $24$ & $\pm~$4 & ${}^{+6}_{-4}$  \\
$  2.50~-~  2.80$~~~  & $56$ & $\pm~$8 & ${}^{+12}_{-9}$  \\
$  2.80~-~  3.00$~~~  & $121$ & $\pm~$13 & ${}^{+24}_{-17}$  \\
$  3.00~-~  3.14$~~~  & $173$ & $\pm~$16 & ${}^{+36}_{-29}$  \\
[4pt]\hline\hline

\end{tabular}

  \end{center}
\end{table}

\begin{table*}
  \begin{center}
    \caption{Breakdown of the total cross-section systematic uncertainty, for
      each bin of \mgg, \ptgg\ and \dphigg.
      The meaning of each column is as follows:
      ``\nontight'' is the definition of the non-\textsc{tight}
      control sample; ``\nonisol'' is the choice of the \Etisol\
      region used to normalize the non-\textsc{tight} sample;
      ``matrix'' refers to the statistical uncertainty of the matrix
      coefficients used by the event weighting; ``$e\rightarrow\gamma$'' is the total systematic
      coming from the electron fake rate; ``ID'' is the overall
      uncertainty coming from the method used to derive the
      identification efficiency; ``material'' is the effect of
      introducing a detector description with distorted material
      distribution; ``generator'' shows the variation due to the usage
      of a different generator (\textsc{Sherpa} instead of
      \textsc{Pythia}); ``$\sigma_E$'' and ``$E$-scale'' are due to
      uncertainties on energy resolution and scale; ``\Etpartisol'' is
      the effect of smearing the particle-level isolation \Etpartisol;
      ``$\int L\de{}t$'' is the effect due to the total luminosity uncertainty.
      Values quoted as $0.000$ are actually less than $0.0005$ in absolute value.
    }
    \label{tab:cross-sections-breakdown}
    \setlength{\extrarowheight}{4pt}
\begin{tabular}{clllllllllll}
\hline\hline
$\mgg~\mathrm{[GeV]}$ & \multicolumn{1}{c}{\nontight} & \multicolumn{1}{c}{\nonisol} & \multicolumn{1}{c}{matrix} & \multicolumn{1}{c}{$e\rightarrow\gamma$} & \multicolumn{1}{c}{ID} & \multicolumn{1}{c}{material} & \multicolumn{1}{c}{generator} & \multicolumn{1}{c}{$\sigma_E$} & \multicolumn{1}{c}{$E$-scale} & \multicolumn{1}{c}{$\Etpartisol$} & \multicolumn{1}{c}{$\int L\de t$} \\
[4pt]\hline
$     0~-~    30$~~~  & ${}^{+0.03}_{-0.01}$ & ${}^{+0.000}_{-0.005}$ & ${}^{+0.021}_{-0.022}$ & ${}^{+0.002}_{-0.002}$ & ${}^{+0.020}_{-0.017}$ & ${}^{+0.021}_{-0.000}$ & ${}^{+0.03}_{-0.00}$ & ${}^{+0.001}_{-0.000}$ & ${}^{+0.006}_{-0.002}$ & ${}^{+0.000}_{-0.010}$ & ${}^{+0.007}_{-0.007}$ \\
$    30~-~    40$~~~  & ${}^{+0.17}_{-0.09}$ & ${}^{+0.00}_{-0.05}$ & ${}^{+0.13}_{-0.13}$ & ${}^{+0.008}_{-0.008}$ & ${}^{+0.22}_{-0.18}$ & ${}^{+0.3}_{-0.0}$ & ${}^{+0.014}_{-0.000}$ & ${}^{+0.003}_{-0.000}$ & ${}^{+0.04}_{-0.03}$ & ${}^{+0.00}_{-0.09}$ & ${}^{+0.06}_{-0.06}$ \\
$    40~-~    50$~~~  & ${}^{+0.3}_{-0.1}$ & ${}^{+0.00}_{-0.06}$ & ${}^{+0.19}_{-0.19}$ & ${}^{+0.008}_{-0.008}$ & ${}^{+0.24}_{-0.20}$ & ${}^{+0.3}_{-0.0}$ & ${}^{+0.11}_{-0.00}$ & ${}^{+0.024}_{-0.000}$ & ${}^{+0.09}_{-0.03}$ & ${}^{+0.00}_{-0.17}$ & ${}^{+0.08}_{-0.08}$ \\
$    50~-~    60$~~~  & ${}^{+0.20}_{-0.13}$ & ${}^{+0.00}_{-0.04}$ & ${}^{+0.14}_{-0.14}$ & ${}^{+0.007}_{-0.007}$ & ${}^{+0.15}_{-0.13}$ & ${}^{+0.19}_{-0.00}$ & ${}^{+0.05}_{-0.00}$ & ${}^{+0.003}_{-0.000}$ & ${}^{+0.06}_{-0.03}$ & ${}^{+0.00}_{-0.13}$ & ${}^{+0.06}_{-0.06}$ \\
$    60~-~    70$~~~  & ${}^{+0.14}_{-0.06}$ & ${}^{+0.001}_{-0.016}$ & ${}^{+0.09}_{-0.09}$ & ${}^{+0.004}_{-0.004}$ & ${}^{+0.05}_{-0.04}$ & ${}^{+0.07}_{-0.00}$ & ${}^{+0.04}_{-0.00}$ & ${}^{+0.007}_{-0.000}$ & ${}^{+0.03}_{-0.02}$ & ${}^{+0.00}_{-0.05}$ & ${}^{+0.03}_{-0.03}$ \\
$    70~-~    80$~~~  & ${}^{+0.06}_{-0.06}$ & ${}^{+0.000}_{-0.007}$ & ${}^{+0.05}_{-0.06}$ & ${}^{+0.003}_{-0.003}$ & ${}^{+0.03}_{-0.03}$ & ${}^{+0.07}_{-0.00}$ & ${}^{+0.03}_{-0.00}$ & ${}^{+0.002}_{-0.001}$ & ${}^{+0.009}_{-0.002}$ & ${}^{+0.00}_{-0.03}$ & ${}^{+0.015}_{-0.015}$ \\
$    80~-~   100$~~~  & ${}^{+0.04}_{-0.05}$ & ${}^{+0.000}_{-0.005}$ & ${}^{+0.04}_{-0.04}$ & ${}^{+0.019}_{-0.019}$ & ${}^{+0.03}_{-0.02}$ & ${}^{+0.04}_{-0.00}$ & ${}^{+0.012}_{-0.000}$ & ${}^{+0.004}_{-0.000}$ & ${}^{+0.013}_{-0.003}$ & ${}^{+0.00}_{-0.03}$ & ${}^{+0.013}_{-0.013}$ \\
$   100~-~   150$~~~  & ${}^{+0.019}_{-0.016}$ & ${}^{+0.001}_{-0.001}$ & ${}^{+0.015}_{-0.018}$ & ${}^{+0.001}_{-0.001}$ & ${}^{+0.004}_{-0.003}$ & ${}^{+0.002}_{-0.001}$ & ${}^{+0.004}_{-0.000}$ & ${}^{+0.000}_{-0.001}$ & ${}^{+0.002}_{-0.003}$ & ${}^{+0.000}_{-0.005}$ & ${}^{+0.003}_{-0.003}$ \\
$   150~-~   200$~~~  & ${}^{+0.002}_{-0.002}$ & ${}^{+0.000}_{-0.000}$ & ${}^{+0.003}_{-0.003}$ & ${}^{+0.000}_{-0.000}$ & ${}^{+0.002}_{-0.001}$ & ${}^{+0.004}_{-0.000}$ & ${}^{+0.001}_{-0.000}$ & ${}^{+0.000}_{-0.000}$ & ${}^{+0.001}_{-0.000}$ & ${}^{+0.000}_{-0.002}$ & ${}^{+0.001}_{-0.001}$ \\
[4pt]\hline\hline
$\ptgg~\mathrm{[GeV]}$ & \multicolumn{1}{c}{\nontight} & \multicolumn{1}{c}{\nonisol} & \multicolumn{1}{c}{matrix} & \multicolumn{1}{c}{$e\rightarrow\gamma$} & \multicolumn{1}{c}{ID} & \multicolumn{1}{c}{material} & \multicolumn{1}{c}{generator} & \multicolumn{1}{c}{$\sigma_E$} & \multicolumn{1}{c}{$E$-scale} & \multicolumn{1}{c}{$\Etpartisol$} & \multicolumn{1}{c}{$\int L\de t$} \\
[4pt]\hline
$     0~-~    10$~~~  & ${}^{+0.3}_{-0.2}$ & ${}^{+0.00}_{-0.09}$ & ${}^{+0.3}_{-0.3}$ & ${}^{+0.03}_{-0.03}$ & ${}^{+0.4}_{-0.4}$ & ${}^{+0.6}_{-0.0}$ & ${}^{+0.10}_{-0.00}$ & ${}^{+0.03}_{-0.00}$ & ${}^{+0.12}_{-0.05}$ & ${}^{+0.0}_{-0.3}$ & ${}^{+0.15}_{-0.15}$ \\
$    10~-~    20$~~~  & ${}^{+0.3}_{-0.2}$ & ${}^{+0.00}_{-0.05}$ & ${}^{+0.21}_{-0.22}$ & ${}^{+0.015}_{-0.015}$ & ${}^{+0.20}_{-0.17}$ & ${}^{+0.21}_{-0.00}$ & ${}^{+0.11}_{-0.00}$ & ${}^{+0.001}_{-0.001}$ & ${}^{+0.06}_{-0.03}$ & ${}^{+0.00}_{-0.15}$ & ${}^{+0.08}_{-0.08}$ \\
$    20~-~    30$~~~  & ${}^{+0.21}_{-0.16}$ & ${}^{+0.000}_{-0.025}$ & ${}^{+0.13}_{-0.14}$ & ${}^{+0.008}_{-0.008}$ & ${}^{+0.07}_{-0.06}$ & ${}^{+0.10}_{-0.00}$ & ${}^{+0.022}_{-0.000}$ & ${}^{+0.010}_{-0.000}$ & ${}^{+0.03}_{-0.02}$ & ${}^{+0.00}_{-0.08}$ & ${}^{+0.03}_{-0.03}$ \\
$    30~-~    40$~~~  & ${}^{+0.13}_{-0.08}$ & ${}^{+0.000}_{-0.012}$ & ${}^{+0.09}_{-0.10}$ & ${}^{+0.006}_{-0.006}$ & ${}^{+0.06}_{-0.05}$ & ${}^{+0.11}_{-0.00}$ & ${}^{+0.08}_{-0.00}$ & ${}^{+0.007}_{-0.000}$ & ${}^{+0.015}_{-0.009}$ & ${}^{+0.00}_{-0.03}$ & ${}^{+0.021}_{-0.021}$ \\
$    40~-~    50$~~~  & ${}^{+0.08}_{-0.06}$ & ${}^{+0.000}_{-0.007}$ & ${}^{+0.05}_{-0.06}$ & ${}^{+0.004}_{-0.004}$ & ${}^{+0.018}_{-0.017}$ & ${}^{+0.03}_{-0.00}$ & ${}^{+0.005}_{-0.000}$ & ${}^{+0.000}_{-0.012}$ & ${}^{+0.00}_{-0.03}$ & ${}^{+0.000}_{-0.015}$ & ${}^{+0.009}_{-0.009}$ \\
$    50~-~    60$~~~  & ${}^{+0.03}_{-0.03}$ & ${}^{+0.000}_{-0.007}$ & ${}^{+0.02}_{-0.03}$ & ${}^{+0.006}_{-0.006}$ & ${}^{+0.03}_{-0.02}$ & ${}^{+0.04}_{-0.00}$ & ${}^{+0.04}_{-0.00}$ & ${}^{+0.013}_{-0.000}$ & ${}^{+0.05}_{-0.01}$ & ${}^{+0.000}_{-0.023}$ & ${}^{+0.012}_{-0.012}$ \\
$    60~-~    80$~~~  & ${}^{+0.021}_{-0.023}$ & ${}^{+0.000}_{-0.001}$ & ${}^{+0.014}_{-0.016}$ & ${}^{+0.001}_{-0.001}$ & ${}^{+0.003}_{-0.003}$ & ${}^{+0.005}_{-0.000}$ & ${}^{+0.000}_{-0.004}$ & ${}^{+0.000}_{-0.001}$ & ${}^{+0.000}_{-0.002}$ & ${}^{+0.000}_{-0.004}$ & ${}^{+0.002}_{-0.002}$ \\
$    80~-~   100$~~~  & ${}^{+0.006}_{-0.000}$ & ${}^{+0.000}_{-0.001}$ & ${}^{+0.005}_{-0.005}$ & ${}^{+0.002}_{-0.002}$ & ${}^{+0.003}_{-0.002}$ & ${}^{+0.002}_{-0.006}$ & ${}^{+0.000}_{-0.005}$ & ${}^{+0.001}_{-0.000}$ & ${}^{+0.004}_{-0.001}$ & ${}^{+0.000}_{-0.004}$ & ${}^{+0.002}_{-0.002}$ \\
$   100~-~   150$~~~  & ${}^{+0.002}_{-0.001}$ & ${}^{+0.000}_{-0.000}$ & ${}^{+0.001}_{-0.002}$ & ${}^{+0.000}_{-0.000}$ & ${}^{+0.000}_{-0.000}$ & ${}^{+0.000}_{-0.000}$ & ${}^{+0.000}_{-0.001}$ & ${}^{+0.000}_{-0.000}$ & ${}^{+0.000}_{-0.000}$ & ${}^{+0.000}_{-0.000}$ & ${}^{+0.000}_{-0.000}$ \\
$   150~-~   200$~~~  & ${}^{+0.000}_{-0.000}$ & ${}^{+0.000}_{-0.000}$ & ${}^{+0.000}_{-0.000}$ & ${}^{+0.000}_{-0.000}$ & ${}^{+0.000}_{-0.000}$ & ${}^{+0.000}_{-0.000}$ & ${}^{+0.000}_{-0.000}$ & ${}^{+0.000}_{-0.000}$ & ${}^{+0.000}_{-0.000}$ & ${}^{+0.000}_{-0.000}$ & ${}^{+0.000}_{-0.000}$ \\
[4pt]\hline\hline
$\dphigg~\mathrm{[rad]}$ & \multicolumn{1}{c}{\nontight} & \multicolumn{1}{c}{\nonisol} & \multicolumn{1}{c}{matrix} & \multicolumn{1}{c}{$e\rightarrow\gamma$} & \multicolumn{1}{c}{ID} & \multicolumn{1}{c}{material} & \multicolumn{1}{c}{generator} & \multicolumn{1}{c}{$\sigma_E$} & \multicolumn{1}{c}{$E$-scale} & \multicolumn{1}{c}{$\Etpartisol$} & \multicolumn{1}{c}{$\int L\de t$} \\
[4pt]\hline
$  0.00~-~  1.00$~~~  & ${}^{+1.1}_{-0.5}$ & ${}^{+0.00}_{-0.14}$ & ${}^{+0.8}_{-0.8}$ & ${}^{+0.05}_{-0.05}$ & ${}^{+0.4}_{-0.4}$ & ${}^{+0.4}_{-0.0}$ & ${}^{+0.3}_{-0.0}$ & ${}^{+0.000}_{-0.017}$ & ${}^{+0.14}_{-0.08}$ & ${}^{+0.0}_{-0.3}$ & ${}^{+0.17}_{-0.17}$ \\
$  1.00~-~  2.00$~~~  & ${}^{+1.6}_{-1.0}$ & ${}^{+0.0}_{-0.3}$ & ${}^{+1.2}_{-1.2}$ & ${}^{+0.07}_{-0.07}$ & ${}^{+0.8}_{-0.7}$ & ${}^{+1.0}_{-0.0}$ & ${}^{+0.5}_{-0.0}$ & ${}^{+0.023}_{-0.000}$ & ${}^{+0.23}_{-0.10}$ & ${}^{+0.0}_{-0.5}$ & ${}^{+0.3}_{-0.3}$ \\
$  2.00~-~  2.50$~~~  & ${}^{+3}_{-2}$ & ${}^{+0.0}_{-0.4}$ & ${}^{+2.2}_{-2.3}$ & ${}^{+0.17}_{-0.17}$ & ${}^{+2.2}_{-1.8}$ & ${}^{+3}_{-0}$ & ${}^{+1.5}_{-0.0}$ & ${}^{+0.10}_{-0.00}$ & ${}^{+0.6}_{-0.4}$ & ${}^{+0.0}_{-1.3}$ & ${}^{+0.8}_{-0.8}$ \\
$  2.50~-~  2.80$~~~  & ${}^{+6}_{-5}$ & ${}^{+0.0}_{-1.3}$ & ${}^{+5}_{-5}$ & ${}^{+0.4}_{-0.4}$ & ${}^{+5}_{-4}$ & ${}^{+6}_{-0}$ & ${}^{+0.3}_{-0.0}$ & ${}^{+0.4}_{-0.0}$ & ${}^{+1.8}_{-1.0}$ & ${}^{+0}_{-4}$ & ${}^{+1.9}_{-1.9}$ \\
$  2.80~-~  3.00$~~~  & ${}^{+11}_{-5}$ & ${}^{+0}_{-3}$ & ${}^{+9}_{-10}$ & ${}^{+0.9}_{-0.9}$ & ${}^{+11}_{-9}$ & ${}^{+14}_{-0}$ & ${}^{+2.3}_{-0.0}$ & ${}^{+0.7}_{-0.0}$ & ${}^{+4}_{-1}$ & ${}^{+0}_{-9}$ & ${}^{+4}_{-4}$ \\
$  3.00~-~  3.14$~~~  & ${}^{+19}_{-16}$ & ${}^{+0}_{-3}$ & ${}^{+14}_{-15}$ & ${}^{+1.5}_{-1.5}$ & ${}^{+16}_{-13}$ & ${}^{+18}_{-0}$ & ${}^{+9}_{-0}$ & ${}^{+0.6}_{-0.0}$ & ${}^{+4}_{-2}$ & ${}^{+0}_{-12}$ & ${}^{+6}_{-6}$ \\
[4pt]\hline\hline

\end{tabular}

  \end{center}
\end{table*}

\section{Conclusions}
\label{sec-9}

This paper describes the measurement of the production cross-section
of isolated di-photon final states in proton-proton collisions, at a
centre-of-mass energy $\sqrt{s}=7$~TeV, with the ATLAS experiment. The
full data sample collected in 2010, corresponding to an
integrated luminosity of $37.2\pm1.3~\ipb$, has been analysed.

The selected sample consists of 2022 candidate events containing two
reconstructed photons, with transverse momenta $\pt>16$~GeV and
satisfying tight identification and isolation requirements.  All the
background sources have been investigated with data-driven techniques
and subtracted.  The main background source, due to hadronic jets in
photon-jet and di-jet events, has been estimated with three
computationally independent analyses, all based on shower shape
variables and isolation, which give compatible results.  The
background due to isolated electrons from $W$ and $Z$ decays is
estimated with collision data, from the proportions of observed
\el\el, \ph\el\ and \ph\ph\ final states, in the $Z$-mass region and
elsewhere.  

The result is presented in terms of differential cross-sections as
functions of three observables: the invariant mass \mgg, the total
transverse momentum \ptgg, and the azimuthal separation \dphigg\ of
the photon pair. The experimental results are compared with NLO
predictions obtained with DIPHOX and ResBos generators.  The observed
spectrum of $\de\sigma/\de\dphigg$ is broader than the NLO
predictions.  The distribution of $\de\sigma/\de\mgg$ is in good
agreement with both the DIPHOX and ResBos predictions, apart from the
low mass region.  The result for $\de\sigma/\de\ptgg$ is generally
well described by DIPHOX and ResBos.

\section{Acknowledgements}

We thank CERN for the very successful operation of the LHC, as well as the
support staff from our institutions without whom ATLAS could not be
operated efficiently.

We acknowledge the support of ANPCyT, Argentina; YerPhI, Armenia; ARC,
Australia; BMWF, Austria; ANAS, Azerbaijan; SSTC, Belarus; CNPq and FAPESP,
Brazil; NSERC, NRC and CFI, Canada; CERN; CONICYT, Chile; CAS, MOST and
NSFC, China; COLCIENCIAS, Colombia; MSMT CR, MPO CR and VSC CR, Czech
Republic; DNRF, DNSRC and Lundbeck Foundation, Denmark; ARTEMIS, European
Union; IN2P3-CNRS, CEA-DSM/IRFU, France; GNAS, Georgia; BMBF, DFG, HGF, MPG
and AvH Foundation, Germany; GSRT, Greece; ISF, MINERVA, GIF, DIP and
Benoziyo Center, Israel; INFN, Italy; MEXT and JSPS, Japan; CNRST, Morocco;
FOM and NWO, Netherlands; RCN, Norway; MNiSW, Poland; GRICES and FCT,
Portugal; MERYS (MECTS), Romania; MES of Russia and ROSATOM, Russian
Federation; JINR; MSTD, Serbia; MSSR, Slovakia; ARRS and MVZT, Slovenia;
DST/NRF, South Africa; MICINN, Spain; SRC and Wallenberg Foundation,
Sweden; SER, SNSF and Cantons of Bern and Geneva, Switzerland; NSC, Taiwan;
TAEK, Turkey; STFC, the Royal Society and Leverhulme Trust, United Kingdom;
DOE and NSF, United States of America.

The crucial computing support from all WLCG partners is acknowledged
gratefully, in particular from CERN and the ATLAS Tier-1 facilities at
TRIUMF (Canada), NDGF (Denmark, Norway, Sweden), CC-IN2P3 (France),
KIT/GridKA (Germany), INFN-CNAF (Italy), NL-T1 (Netherlands), PIC (Spain),
ASGC (Taiwan), RAL (UK) and BNL (USA) and in the Tier-2 facilities
worldwide.

\bibliography{DiphotonCrossSection}
\onecolumngrid

\clearpage
\onecolumngrid \
\begin{flushleft}
{\Large The ATLAS Collaboration}

\bigskip

G.~Aad$^{\rm 48}$,
B.~Abbott$^{\rm 111}$,
J.~Abdallah$^{\rm 11}$,
A.A.~Abdelalim$^{\rm 49}$,
A.~Abdesselam$^{\rm 118}$,
O.~Abdinov$^{\rm 10}$,
B.~Abi$^{\rm 112}$,
M.~Abolins$^{\rm 88}$,
H.~Abramowicz$^{\rm 153}$,
H.~Abreu$^{\rm 115}$
% ...
\bigskip

% ...
$^{10}$ Institute of Physics, Azerbaijan Academy of Sciences, Baku, Azerbaijan\\
$^{11}$ Institut de F\'isica d'Altes Energies and Universitat Aut\`onoma  de Barcelona and ICREA, Barcelona, Spain\\
% ...
$^{48}$ Fakult\"{a}t f\"{u}r Mathematik und Physik, Albert-Ludwigs-Universit\"{a}t, Freiburg i.Br., Germany\\
$^{49}$ Section de Physique, Universit\'e de Gen\`eve, Geneva, Switzerland\\
% ...
$^{88}$ Department of Physics and Astronomy, Michigan State University, East Lansing MI, United States of America\\
% ...
$^{111}$ Homer L. Dodge Department of Physics and Astronomy, University of Oklahoma, Norman OK, United States of America\\
$^{112}$ Department of Physics, Oklahoma State University, Stillwater OK, United States of America\\
% ...
$^{115}$ LAL, Univ. Paris-Sud and CNRS/IN2P3, Orsay, France\\
% ...
$^{118}$ Department of Physics, Oxford University, Oxford, United Kingdom\\
% ...
$^{153}$ Raymond and Beverly Sackler School of Physics and Astronomy, Tel Aviv University, Tel Aviv, Israel\\
% ...
\end{flushleft}


\end{document}